\documentclass[pra,superscriptaddress,twocolumn]{revtex4}

\usepackage{enumerate}
\usepackage{amsfonts,amssymb,amsmath}
\usepackage[]{graphics,graphicx,epsfig}
\usepackage{amsthm}
\usepackage{float}

\usepackage{graphicx}
\usepackage{dcolumn}
\usepackage{natbib}
\usepackage{color}
\usepackage{multirow}
\usepackage{ulem,xpatch}

\def\identity{\leavevmode\hbox{\small1\kern-3.8pt\normalsize1}}

\newtheorem{propo}{Proposition}
\newcommand{\be}{\begin{eqnarray}}
\newcommand{\ee}{\end{eqnarray}}
\newcommand{\bpr}{\begin{propo}}
\newcommand{\epr}{\end{propo}}
\newcommand{\bpf}{\begin{proof}}
\newcommand{\epf}{\end{proof}}
\newcommand{\ket}[1]{\left | #1 \right\rangle}
\newcommand{\bra}[1]{\left \langle #1 \right |}

\renewcommand{\epsilon}{\varepsilon}

\begin{document}


\title{Assembly of 2N entangled fermions into multipartite composite bosons}

\author{Zakarya Lasmar}   
\affiliation{Faculty of Physics, Adam Mickiewicz University, Umultowska 85, 61-614 Pozna\'n, Poland}

\author{P. Alexander Bouvrie}   
\affiliation{Centro Brasileiro de Pesquisas Fisicas, Rua Dr. Xavier Sigaud 150, Rio de Janeiro, RJ 22290-180, Brazil}

\author{Adam S. Sajna}   
\affiliation{Faculty of Physics, Adam Mickiewicz University, Umultowska 85, 61-614 Pozna\'n, Poland}

\author{Malte C. Tichy}   
\affiliation{Department of Physics and Astronomy, University of Aarhus, DK–8000 Aarhus C, Denmark}

\author{Pawe\l{} Kurzy\'nski}   \email{pawel.kurzynski@amu.edu.pl}   
\affiliation{Faculty of Physics, Adam Mickiewicz University, Umultowska 85, 61-614 Pozna\'n, Poland}
\affiliation{Centre for Quantum Technologies, National University of Singapore, 3 Science Drive 2, 117543 Singapore, Singapore}

\date{\today}


\begin{abstract}

An even number of fermions can behave in a bosonic way. The simplest scenario involves two fermions which can form a single boson. But four fermions can either behave as two bipartite bosons or further assemble into a single four-partite bosonic molecule. In general, for 2N fermions there are many possible arrangements into composite bosons. The question is: what determines which fermionic arrangement is going to be realized in a given situation and can such arrangement be considered truly bosonic? This work aims to find the answer to the above question. We propose an entanglement-based method to assess bosonic quality of fermionic arrangements and apply it to study how the ground state of the extended one-dimensional Hubbard model changes as the strength of intra-particle interactions increases. 

\end{abstract}

\maketitle


\section{Introduction}

Most bosons studied in laboratories are in fact composed of elementary fermions. It is therefore important to understand what makes them behave in a bosonic way. One may think that it is the binding force that keeps them together, however recently an intriguing idea was proposed that the bosonic behaviour stems from the intra-fermionic entanglement. More precisely, the more entangled two fermions are, the more bosonic they behave \cite{Law,Wootters}. This result established a new field of research, the quantum information oriented studies on composite bosons \cite{pawel2011,Bouvrie2012a,Bouvrie2012b,Bouvrie2013,Bouvrie2014,Bouvrie2016,Bouvrie2017,Bouvrie2018,Pawel2012,Pawel2013,Pawel2015,zak2017,zak2018}. Up to now scientists focussed on composite bosons made of two elementary components, whose quality can be quantified by bipartite entanglement. Here, we propose a generalization to composite bosons made of $2N$ fermions, whose quality is described by genuine multipartite entanglement. 

The main problem of this work can be formulated in the following way. Suppose that $2N$ fermions are in the state 
\begin{equation}
|\psi\rangle=\sum_{i_1,i_2,\ldots,i_{2N}}\alpha_{i_1,i_2,\ldots,i_{2N}} a^{\dagger}_{i_1}a^{\dagger}_{i_2}\ldots a^{\dagger}_{i_{2N}}|0\rangle,
\end{equation}
where $a^{\dagger}_{i_k}$ creates a fermion in the mode $i_k$ and $\{\alpha_{i_1,i_2,\ldots,i_{2N}}\}$ is the set of antisymmetric coefficients. We ask: does the state $|\psi\rangle$ describe a single composite bosonic particle made of $2N$ fermions? Or perhaps it describes two bosonic particles, each made of $N$ fermions? Or maybe it describes $N$ bipartite bosonic particles? In fact, for $2N$ fermions there is a number of possible bosonic assemblies. How to decide to which assembly the state $|\psi\rangle$ corresponds to and how to quantify its bosonic quality? We are going to address the above questions. In particular, we will focus on fermionic states that are ground states of the one-dimensional Hubbard model. We will study how assemblies of $2N$ spin-$1/2$ particles on a lattice, and their corresponding bosonic qualities, depend on the strength of intra-particle interactions.


Apart from contributing to the new field of research, there are two additional motivations behind our studies. The first one is fundamental and is related to general investigations on complexity in the quantum domain. In particular, we want to understand what makes a complex quantum system to behave as a single entity. The second one is more pragmatic. We would like to understand how to engineer and control the creation of composite particles. This is related to the following problem -- in some situations spontaneous emergence of composite structures can affect the property one wants to observe. A very illustrative example is the problem of the Bose-Einstein condensation of atomic Hydrogen \cite{Fried1998}. Hydrogen atoms naturally try to recombine into $H_2$ molecules and one needs to find out a method to prevent it (for example by spin polarization in high magnetic fields \cite{Silvera1980}). Here, we consider a system of $2N$ interacting fermions on a one-dimensional lattice and show that the formation of composite structures can be controlled by a proper tuning of the nearest neighbour interaction.


\section{Bipartite composite bosons}

In this section we recall properties of composite bosons made of two fermions and show that such composite bosons naturally describe ground states of the Hubbard model for two fermionic particles.   

\subsection{Bipartite composite bosons and the role of correlations}

Consider a system made of two fermions whose general state is given by
\begin{equation}
|\psi\rangle = \sum_{i,j} \alpha_{i,j} a_i^{\dagger} a_j^{\dagger} |0\rangle,
\end{equation} 
where the matrix of coefficients $\alpha_{i,j}$ satisfies $\alpha_{i,j}=-\alpha_{j,i}$, $\alpha_{i,i}=0$ and $\sum_{i,j}|\alpha_{i,j}|^2 = 1$. The above bi-fermionic state admits a Schmidt-like (Slater) representation \cite{Slater1,Slater2,Slater3,Eckert2002}, i.e., there exist a unitary transformation $Ua^{\dagger}_i = \sum_k \beta_{i,k}a^{\dagger}_k$, such that
\begin{equation}
|\varphi\rangle = \sum_{i,j} \alpha_{i,j} U a_i^{\dagger} U a_j^{\dagger} |0\rangle = \sum_{k}\sqrt{\lambda_k}a^{\dagger}_{2k}a^{\dagger}_{2k+1}.
\end{equation}
The above representation allows to divide the fermions into two different groups, the ones occupying even modes and the ones occupying odd modes. We therefore set $a^{\dagger}_{2k} \equiv a^{\dagger}_k$ and $a^{\dagger}_{2k+1} \equiv b^{\dagger}_k$ and assume $k=0,1,\ldots,d-1$, which allows us to write  
\begin{equation}\label{bi-fermion}
|\varphi\rangle = \sum_{k=0}^{d-1} \sqrt{\lambda_k} a_k^{\dagger} b_k^{\dagger} |0\rangle.
\end{equation}  
In many situations the division into the two groups of fermions, call them $A$ and $B$, is quite natural from the physical point of view. For example, $a^{\dagger}_k$ can correspond to creation of a particle with spin up and $b^{\dagger}_k$ to creation of a particle with spin down. In the remaining part of this work we assume that we deal with these two types of fermions and we explicitly use the operators $a_k^{\dagger}$ and $b_k^{\dagger}$.

The real non-negative coefficients $\{\lambda_k\}$ determine the correlations between the two fermions. More precisely, one can define the purity $\frac{1}{d} \leq P \leq 1$ of the system as
\begin{equation}
P=\sum_{k}\lambda_k^2.
\end{equation} 
If $P=1$ the system is separable and in any other case the system is entangled, the smaller the purity the more entangled it is. For a pair of $d$-level systems the smallest purity, corresponding to a maximally entangled state, is $1/d$.

While it is true that an even number of fermions can behave like a boson, the exact conditions under which these fermions can be treated as a single bosonic particle has been studied for a long time \cite{Monique2001,Monique2003a,Monique2003b,Monique2009,Monique2010,Monique2011a,Monique2011b,Thilagam2013,Thilagam2015,Monique2015,Monique2016}, for a review cf. \cite{Monique2008}. An interesting contribution was done recently by \cite{Law}, who showed that the bi-fermionic state (\ref{bi-fermion}) has properties of a single bosonic particle in the limit $P\rightarrow 0$. This result established a connection between the theory of entanglement and the studies on composite particles. The keystone is the idea that the bosonic behaviour of composite system does not stem exactly from interactions, but from entanglement. The above paradigm shift, from interactions to entanglement, is quite subtle. Nevertheless, although entanglement cannot be created without interactions, in principle it does not require interactions to last, once it is created. This allows to study the concept of composite particles that are {\it bound} solely by entanglement \cite{zak2017,zak2018}.  

The idea of Law is based on the ladder structure of bosonic operators. If the state (\ref{bi-fermion}) is to be treated like a state of a single boson
\begin{equation}
\sum_{k=0}^{d-1} \sqrt{\lambda_k} a_k^{\dagger} b_k^{\dagger} |0\rangle \equiv c^{\dagger}|0\rangle = |1\rangle,
\end{equation}
then the creation operator $c^{\dagger}$ should obey 
\begin{equation}
\langle 0|c^{N}c^{\dagger N}|0\rangle = N!.
\end{equation}
However, due to the fact that $c^{\dagger}$ is not a perfect bosonic operator one gets
\begin{equation}
\langle 0|c^{N}c^{\dagger N}|0\rangle = \chi_N N!,
\end{equation}
where $\chi_N$ is a factor describing a departure from perfect bosonic behaviour. In addition 
\begin{equation}
\label{cobState}
|N\rangle \equiv \frac{c^{\dagger N}}{\sqrt{\chi_N N!}}|0\rangle
\end{equation}
and
\begin{eqnarray}
\label{ladder}
c^{\dagger}|N-1\rangle &=& \alpha_N\sqrt{N}|N\rangle, \nonumber \\
c |N\rangle &=& \alpha_N \sqrt{N}|N-1\rangle + |\epsilon_N\rangle,
\end{eqnarray}
where
\begin{equation}
\alpha_N=\sqrt{\frac{\chi_N}{\chi_{N-1}}}
\end{equation}
and $|\epsilon_N\rangle$ is a state of $N-1$ fermionic A-B pairs. This state is orthogonal to the Fock state $|N-1\rangle$. Its norm is
\begin{equation}
\langle \epsilon_N|\epsilon_N \rangle = 1 - N\frac{\chi_N}{\chi_{N-1}}+(N-1)\frac{\chi_{N+1}}{\chi_{N}}.
\end{equation}

Law observed that in the limit $P\rightarrow 0$ one has $\frac{\chi_N}{\chi_{N-1}} \rightarrow 1$ for all $N$. In this case $\alpha_N \rightarrow 1$ and $\langle \epsilon_N|\epsilon_N \rangle \rightarrow 0$, therefore the composite boson operator $c^{\dagger}$ becomes perfect bosonic. Moreover, Law showed that $\frac{\chi_N}{\chi_{N-1}} \leq 1-P$. This result was further strengthen by Chudzicki, Oke and Wootters \cite{Wootters}, who provided the following bounds
\begin{equation}\label{bounds}
1-NP \leq \frac{\chi_N}{\chi_{N-1}} \leq 1-P,
\end{equation}
and by Tichy, Bouvrie and M\o{}lmer who provided the tights bounds for a given $P$ \cite{Bouvrie2012a}. As a result, in the limit $P \ll 1/N$ the composite creation operators become bosonic. 

Interestingly, $P$ is directly related to $\chi_2$
\begin{equation}
\chi_2 = 2\sum_{i<j} \lambda_i \lambda_j =\sum_{i,j} \lambda_i \lambda_j - \sum_{i} \lambda_i^2 =1 - P.
\end{equation}
Therefore, because of bounds (\ref{bounds}), the properties of two composite bosons tell us about properties of more than two such particles. 


\subsection{Maximally entangled bi-fermions}

In our recent work \cite{zak2017} we introduced the concept of a maximally entangled composite boson
\begin{equation}\label{maxentc}
c^{\dagger}|0\rangle = \frac{1}{\sqrt{d}}\sum_{k=0}^{d-1} a_k^{\dagger}b_k^{\dagger}|0\rangle,
\end{equation}
i.e., $\lambda_k=1/d$ for all $k$. The above state has a few nice properties. First of all, the internal structure is described by only one integer $d$. In addition, it is easy to evaluate 
\begin{equation}\label{chi}
\chi_N = \frac{d!}{d^N(d-N)!}
\end{equation}  
and as a result one gets
\begin{equation}
\alpha_N = \sqrt{\frac{d-N+1}{d}},~~~~\langle \epsilon_N|\epsilon_N \rangle = 0.
\end{equation}

Next, let us define the following bi-fermionic states
\begin{equation}\label{bfbasis}
c^{\dagger}_{s,r}|0\rangle = \frac{1}{\sqrt{d}}\sum_{k=0}^{d-1} e^{i\frac{2\pi}{d}kr}a_k^{\dagger}b_{k+s}^{\dagger}|0\rangle,
\end{equation}
where $r,s=0,1,\ldots,d-1$. Each of these states corresponds to a good composite boson for sufficiently large $d$, since for each of these states $P=1/d$. This is the smallest possible $P$ that can be achieved for a bi-partite system with $d$ modes, therefore the creation operators $c^{\dagger}_{s,r}$ provide the best possible bosonic quality. The $d^2$ states form an orthonormal basis because
\begin{equation}
\langle 0|c_{s,r}c^{\dagger}_{s',r'} |0\rangle = \delta_{s,s'}\delta_{r,r'},
\end{equation}
where $\delta_{x,y}$ is the Kronceker delta. Therefore, any bi-fermionic state can be represented as a linear combination of these states. Interestingly, the above orthogonality relation does not imply orthogonality of states corresponding to more than one composite boson. In general $c^{\dagger N}_{s,r} |0\rangle$ and $c^{\dagger N}_{s',r'} |0\rangle$ are not orthogonal. The most extreme case corresponds to $N=d$, for which $c^{\dagger d}_{s,r} |0\rangle \equiv c^{\dagger d}_{s',r'} |0\rangle$ for any $s$, $s'$, $r$ and $r'$. In simple words, for $N=d$ all available modes are filled with fermions and there is exactly one state describing this possibility. The fact that higher power of different composite boson operators do not preserve orthogonality has been already noticed in \cite{Monique2008}. 

Finally, notice that although a perfect boson remains a boson even if it is in a superposition, the superposition of composite boson states
\begin{equation}
\sum_{s,r=0}^{d-1} \alpha_{s,r} c^{\dagger}_{s,r}|0\rangle
\end{equation}
may not correspond to a good composite boson. In the worst case scenario this superposition can correspond to a product state $a^{\dagger}_k b^{\dagger}_{k'}$. However, if the number of terms in superposition is large, the system should still manifest good bosonic behaviour. Still, good bosonic behaviour is always a question of how many composite bosons occupy the same state.


\subsection{Bosonic quality of the Hubbard ground state}

In the limit of strong interactions one can use the hard-core boson approximation (see the end of this section) to prove that the ground state of the Hubbard model is a composite boson state of the form \eqref{bi-fermion} with $\lambda_k = 1/d$. In general, the one dimensional Hubbard model can be solved exactly using Bethe ansatz \cite{Lieb1968}. Here, we show that the bipartite ground state can be found using the above maximally entangled states \eqref{bfbasis}. 

The Hubbard Hamiltonian for our system is given by
\begin{equation}\label{H1}
\mathcal{H} = J\mathcal{H}_0 + U\mathcal{H}_{p} 
\end{equation}
where the parameters $J,U \geq 0$,
\begin{equation}
\mathcal{H}_0 = - \sum_{i = 0}^{d-1} \left(a_{i}^{\dagger} a_{i + 1}^{} + b_{i}^{\dagger} b_{i + 1}^{} + h.c.\right)
\end{equation}
is the kinetic energy (hopping) term and
\begin{equation}
\mathcal{H}_p = - \sum_{i = 0}^{d-1}  a_{i}^{\dagger} a_{i}^{} b_{i}^{\dagger} b_{i}^{}
\end{equation}
describes the attractive point interaction between fermions A and B. In the above formula we assume the periodic boundary conditions $d \equiv 0$.

Next, consider the action of the Hamiltonian (\ref{H1}) on $|s,r\rangle \equiv c^{\dagger}_{s,r}|0\rangle$
\begin{eqnarray}
\mathcal{H}|s,r\rangle =  &-& J(1+e^{i\frac{2\pi}{d}r})|s+1,r\rangle \nonumber \\ 
&-& J(1+e^{-i\frac{2\pi}{d}r})|s-1,r\rangle  \nonumber \\
&-& U\delta_{s,0} |s,r\rangle.
\end{eqnarray}
The Hamiltonian does not change the parameter $r$, therefore for each $r$ one can consider a separate decoupled set of equations. To find the ground state, we need to choose $r$ which minimizes the energy. The hopping amplitude is $-J(1+e^{\pm i\frac{2\pi}{d}r})$ and the greatest negative contribution occurs for $r=0$. Therefore, we fix $r=0$ so the hopping term becomes $-2J$.

In addition, we assume for the moment that the Hamiltonian describes particles hopping on infinite discrete line. Therefore, we get $ - \infty \le s \le \infty$. This corresponds to $d\rightarrow \infty$. As a result, the states (\ref{bfbasis}) have an infinite number of terms and the corresponding purity is 
\begin{equation}
P=\lim_{d\rightarrow \infty} \frac{1}{d} = 0,
\end{equation}
therefore they can be considered perfect bosonic for any number of composite bosons $N$.

We represent the candidate ground state as 
\begin{equation}\label{gs}
|\psi_0\rangle = \sum_{s=-\infty}^{\infty} \alpha_s |s,0\rangle,
\end{equation}
for which
\begin{equation}
\mathcal{H}|\psi_0\rangle = \epsilon |\psi_0\rangle,
\end{equation}
where $\epsilon$ is the ground state energy. The goal is to find the coefficients $\alpha_s$ and the energy $\epsilon$. The corresponding set of recurrence equations consists of typical equations
\begin{equation}
-\frac{\epsilon}{2J}\alpha_s = \alpha_{s+1} + \alpha_{s-1}, \label{as}
\end{equation}
which apply to cases $s\neq 0$, and an atypical equation
\begin{equation}
-\frac{(U+\epsilon)}{2J}\alpha_0 = \alpha_1 + \alpha_{-1}. \label{a0}
\end{equation}
The solution to the above equations (provided in Appendix A) yields 
\begin{equation}
\alpha_s = A\left(\frac{\sqrt{U^2+16J}-U}{4J}\right)^{|s|},
\end{equation}
where $A$ can be determined from normalization. The corresponding energy is
\begin{equation}\label{energy}
\epsilon = -\sqrt{U^2+16J^2}.
\end{equation}

In general, the ground state (\ref{gs}) is a superposition of many bi-fermionic maximally entangled states (\ref{bfbasis}), therefore it does not need to describe a perfect composite boson. However, in the limit $U \gg J$ the energy becomes $\epsilon \rightarrow -U$, $\alpha_s \rightarrow 0$ for $s\neq 0$ and $\alpha_0 \rightarrow 1$. Therefore, in this case the ground state of $\mathcal{H}$ is dominated by $c^{\dagger}_{0,0}|0\rangle$, i.e., it can be considered a perfect composite boson.

The system allows us to address the difference between entanglement and interactions. If the bosonic behaviour of two fermions were determined solely by interaction, one could choose $J=0$ and the Hamiltonian would consist only of the interaction part. The corresponding ground state would be degenerated and would be of the form $\eta^{\dagger}_k|0\rangle \equiv a^{\dagger}_k b^{\dagger}_k|0\rangle$, i.e., any pair of fermions A and B occupying the same mode would be considered a ground state. The operator $\eta_k^{\dagger}$ obeys the following commutation relations 
\begin{equation}\label{hardcore}
\eta_k^{\dagger 2}=0, ~~~~\eta^{\dagger}_{k}\eta^{\dagger}_{k'}=\eta^{\dagger}_{k'}\eta^{\dagger}_{k}.
\end{equation} 
These relations are not bosonic. The A-B pairs generated by such operators are sometimes called hard-core bosons. Note, that $\eta_k^{\dagger}$ creates two fermions in the product state. This is a clear manifestation of the fact that bosonic behaviour needs fermionic entanglement. This entanglement is provided by the introduction of the kinetic energy term which lifts of the degeneracy.


\section{Four-partite composite bosons}


\begin{figure}[t]
\includegraphics[width=0.17 \textwidth]{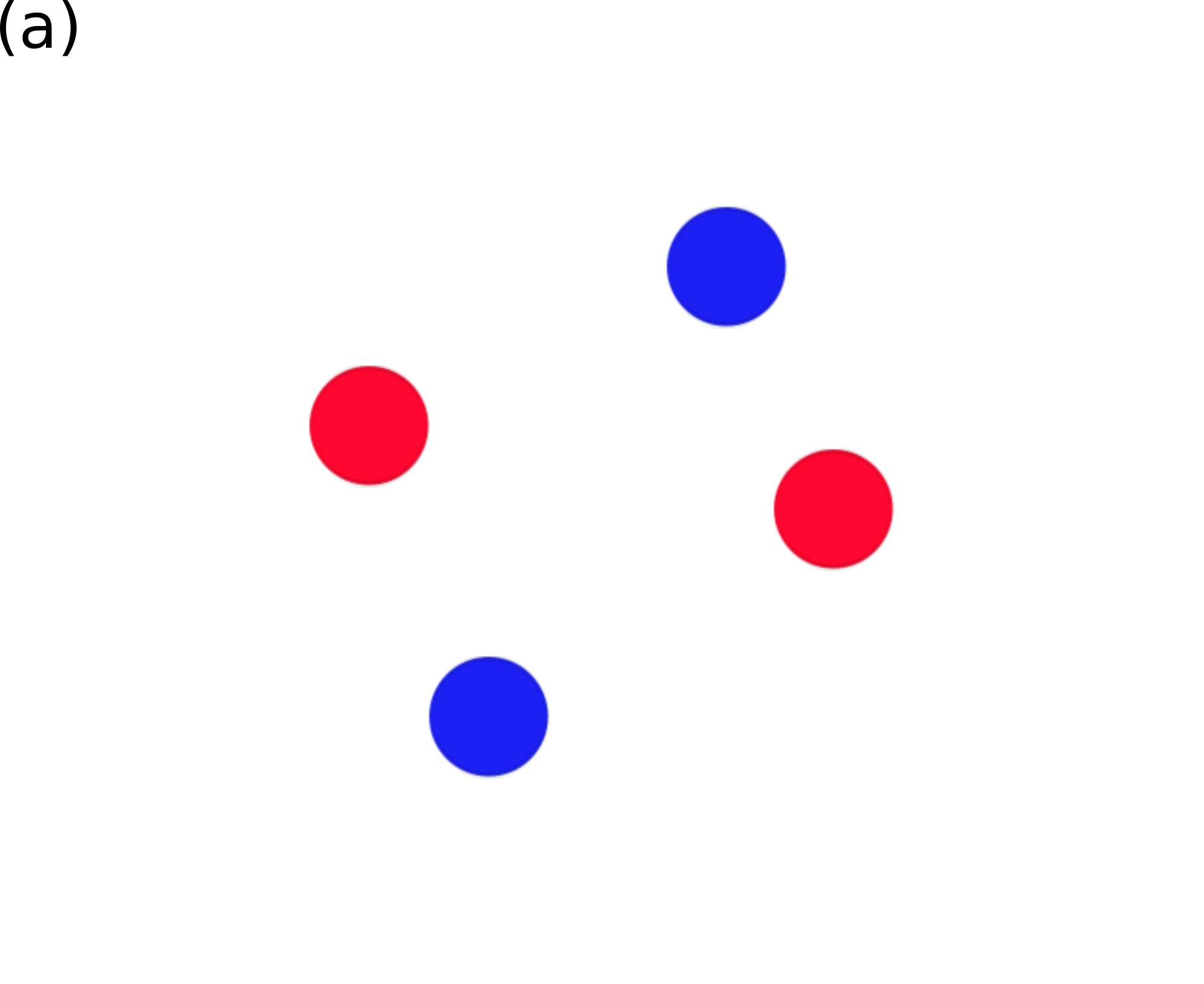}\includegraphics[width=0.17 \textwidth]{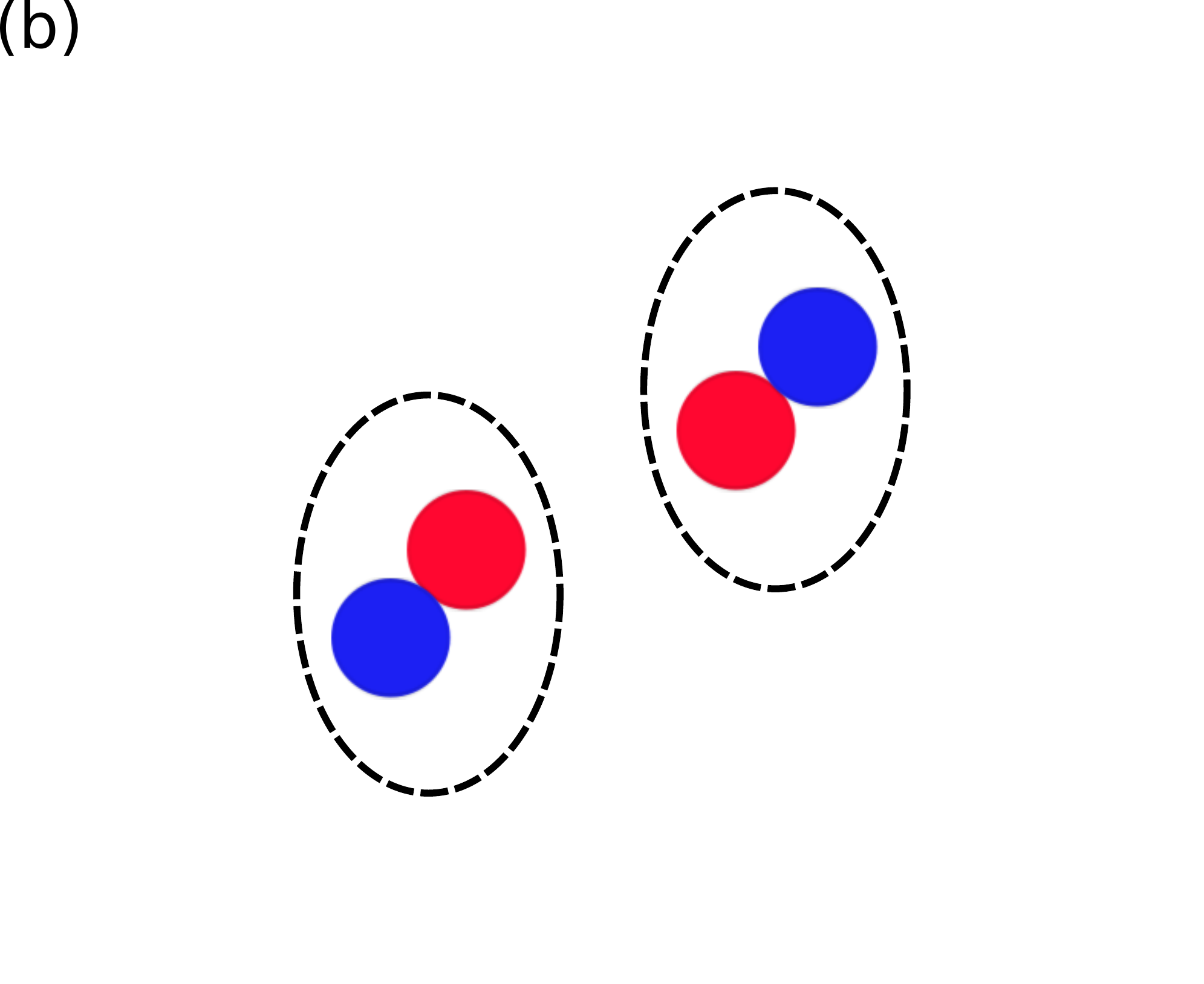}\includegraphics[width=0.17 \textwidth]{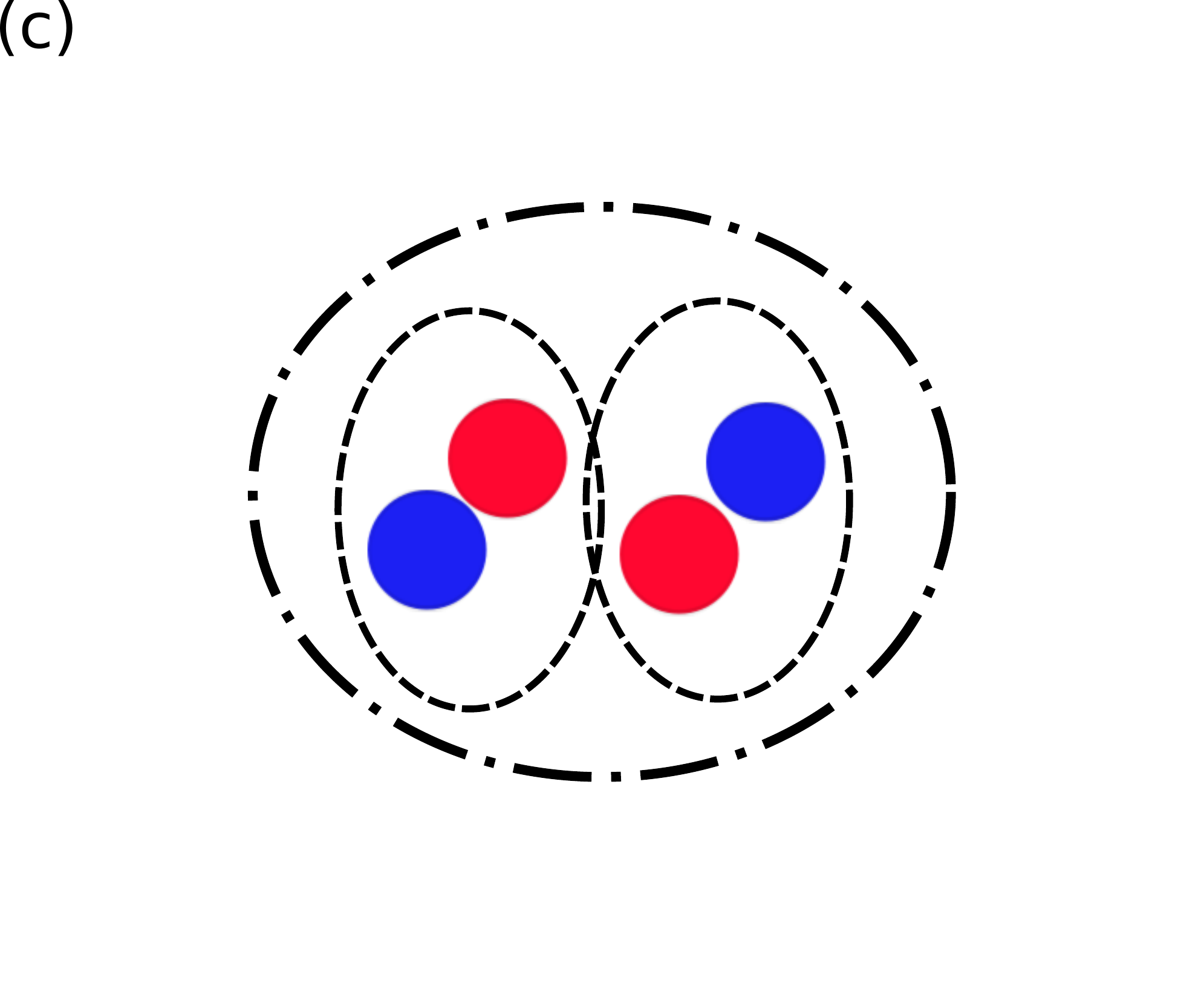}
\caption{Schematic representation of four-fermion assemblies. (a) Free fermions if there are no interactions -- two fermions of type A and two of type B. (b) In the presence of strong point interactions two bosonic A-B pairs emerge. (c) In the presentce of strong point interaction and nearest neighbour interactions a single four-partite bosonic molecule is formed.  \label{fig1}}
\end{figure}

In this section we consider composite bosons made of four fermions. As shown above, the Hubbard Hamiltonian with point interaction can lead to formation of bipartite composite bosons of the A-B type. Now, we would like to observe formation of a single four-partite bosonic molecule (see Fig. \ref{fig1}). Due to Pauli exclusion the bound pair described by $ \eta^{\dagger}_k \equiv a^{\dagger}_k b^{\dagger}_k$ cannot interact via point interaction with any other such pair. Therefore, in order to observe formation of larger compounds we need to introduce a nearest neighbour interaction. 

In the following subsections we introduce a four-partite molecular state and compare it with a state of two bi-partite composite bosons. Next, we analyse which of these states dominate the ground state of the extended Hubbard Hamiltonian as the strength of the nearest neighbour interaction changes. Finally, we analyse bosonic properties of the four-partite molecular state. 


\subsection{Two bipartite composite bosons}

The bipartite ground state of (\ref{H1}) in the limit $U\gg J$ can be approximated as $c^{\dagger}_{0,0}|0\rangle$. Let us represent it in terms of operators $\eta^{\dagger}_k$
\begin{equation}
c^{\dagger}_{0,0}|0\rangle = \frac{1}{\sqrt{d}}\sum_{k=0}^{d-1}\eta^{\dagger}_k|0\rangle,
\end{equation}
From now on we use the operators $\eta^{\dagger}_k$ whenever it is helpful to simplify our notation.

Next, let us consider Fock state representing two such composite bosons
\begin{equation}\label{2bipartite}
\frac{c^{\dagger 2}_{0,0}}{\sqrt{2\chi_2}}|0\rangle = \frac{1}{d\sqrt{2\chi_2}}\sum_{k,k'=0}^{d-1} \eta_k^{\dagger}\eta_{k'}^{\dagger}|0\rangle ,
\end{equation} 
where $\chi_2=1-P=(d-1)/d$. We are going to show that for sufficiently weak nearest neighbour interactions and for $U\gg J$ the above state dominates the ground state of the extended one-dimensional Hubbard model. 


\subsection{Four-partite entangled states}

Consider the following states of four fermions represented via two $\eta^{\dagger}_k$ operators
\begin{equation}\label{fourpartite}
q^{\dagger}_{s,r}|0\rangle \equiv \frac{1}{\sqrt{d}}\sum_{k=0}^{d-1}e^{i\frac{2\pi}{d}kr}\eta_{k}^{\dagger}\eta_{k+ s}^{\dagger}|0\rangle,
\end{equation}
where $r=0,1,\ldots,d - 1$ and $s=1,\ldots,d/2$. The latter results from $\eta_{k}^{\dagger}\eta_{k+ s}^{\dagger}=\eta_{k+ s}^{\dagger}\eta_{k}^{\dagger}$ and we assume that $d$ is even. 

Just like (\ref{bfbasis}), the above states form an orthonormal basis
\begin{equation}
\langle 0|q_{s',r'}q^{\dagger}_{s,r}|0\rangle = \delta_{s,s'}\delta_{r,r'},
\end{equation}
therefore they can be used to represent any state of two A-B pairs created by $\eta_k^{\dagger}$. In particular, the state (\ref{2bipartite}) can be represented as
\begin{equation}\label{biasfour}
\frac{c^{\dagger 2}_{0,0}}{\sqrt{2\chi_2}}|0\rangle = \sqrt{\frac{2}{d}}\sum_{s=1}^{d/2} q^{\dagger}_{s,0}|0\rangle. 
\end{equation}

Note, that the structure of (\ref{fourpartite}) differs from the one of (\ref{2bipartite}). Due to correlations between A-B pairs the number of terms in the first one is quadratically smaller than in the second one. Writing these states explicitly using operators $a^{\dagger}_k$ and $b^{\dagger}_k$ we see that terms in (\ref{fourpartite}) are of the form $a^{\dagger}_k b^{\dagger}_k a^{\dagger}_{k+s} b^{\dagger}_{k+s}$, which indicates four-partite correlations since $s$ is a constant. On the other hand, terms in (\ref{2bipartite}) are of the form $a^{\dagger}_k b^{\dagger}_k a^{\dagger}_{k'} b^{\dagger}_{k'}$, which indicates bipartite correlations since $k$ and $k'$ are independent indices. We already know that bipartite entangled fermionic states exhibit bosonic properties. At the end of this section we are going to examine bosonic properties of four-partite entangled states.


\subsection{Extended Hubbard model}

In order to study the formation of four-partite composite bosons we extend the Hamiltonian (\ref{H1}) to include the nearest neighbour interaction. The new Hamiltonian is of the form
\begin{equation}\label{H}
\mathcal{H} = J\mathcal{H}_0 + U\mathcal{H}_{p} + \gamma \mathcal{H}_{nn} 
\end{equation}
where $J,U,\gamma \geq 0$ and the new term is
\begin{equation}
\mathcal{H}_{nn} =  - \sum_{k = 0}^{d-1}  a_{k}^{\dagger} a_{k}^{}  b_{k+1}^{\dagger} b_{k+1}^{}
\end{equation}
It represents an attractive nearest neighbour interaction between fermions A and B. This is a simplified model, since we could also introduce an attractive nearest neighbour interaction between fermions A and A or B and B. However, because in our model A interacts with B via point interaction, the nearest neighbour interaction between the fermions of the same type is going to be mediated by the fermions of the other type. For example, the fermion A in mode $k$ is going to interact with fermion B in mode $k$ that interacts with another fermion A in mode $k+1$. This leads to indirect interactions between A in mode $k$ and A in mode $k+1$.  Moreover, we are going to show below that in the particularly interesting limit $U\gg J\gg \gamma$ we can focus on the effective Hamiltonian. The form of this Hamiltonian does not depend on whether we choose interaction between A and B, A and A, or B and B.

In general we need to consider a four-partite problem, however in the limit $U\gg J\gg \gamma$ the above Hamiltonian can be represented in an effective form using operators $\eta^{\dagger}_k$ (for details see Appendix B) 
\begin{eqnarray}
\mathcal{H}_\text{eff} &=& - \frac{2 J^2}{U} \sum_{k=0}^{d-1} ( \eta_k^\dagger  \eta_{k+1} + h.c.) \nonumber \\ &-& \left (2\gamma -\frac{4 J^2}{U} \right) \sum_{k=0}^{d-1}  \eta_k^\dagger  \eta_k \eta_{k+1}^\dagger  \eta_{k+1}, \label{Heff}
\end{eqnarray}
for which our four-partite problem reduces to a bipartite problem.

In order to find the ground state, let us consider the action of (\ref{Heff}) on $|s,r\rangle_q \equiv q^{\dagger}_{s,r}|0\rangle$
\begin{eqnarray}
\mathcal{H}_{eff}|s,r\rangle_q =  &-& \bar{J}(1+e^{i\frac{2\pi}{d}r})|s+1,r\rangle_q \nonumber \\ 
&-& \bar{J}(1+e^{-i\frac{2\pi}{d}r})|s-1,r\rangle_q  \nonumber \\
&-& \bar{\gamma}\delta_{s,1} |s,r\rangle_q,
\end{eqnarray}
where $\bar{J}=2 \frac{J^{2}}{U}$ and $\bar{\gamma}=2(\gamma-\bar{J})$. This resembles the previous case. As before, the parameter $r$ is not affected by (\ref{Heff}), therefore we choose $r=0$ to minimize the kinetic energy. Moreover, we consider $d \rightarrow \infty$, therefore there is no upper bound on $s$.

We assume that the ground state is of the form
\begin{equation}
|\psi_0\rangle = \sum_{s=1}^{\infty}\beta_s |s,0\rangle_q,
\end{equation}
and $\mathcal{H}_{eff}|\psi_0\rangle = \bar{\epsilon}|\psi_0\rangle$
hence we obtain the following set of typical recurrence equations
\begin{equation}
-\frac{\bar{\epsilon}}{2\bar{J}}\beta_s = \beta_{s+1} + \beta_{s-1}, \label{bs}
\end{equation}
which apply to cases $s > 1$, and an atypical equation
\begin{equation}
-\frac{(\bar{\gamma}+\bar{\epsilon})}{2\bar{J}}\beta_1 = \beta_2. \label{b1}
\end{equation}
Using the same methods as before (see Appendix C) we find that 
\begin{eqnarray}
\beta_s &=& B \left(\frac{\bar{J}}{\gamma-\bar{J}}\right)^s, \\
\bar{\epsilon}  &=& \frac{4\gamma\bar{J}-4\bar{J}^2-2\gamma^2}{\gamma-\bar{J}}, 
\end{eqnarray}
where $B$ is the normalization constant. The above solution works for $U \gg \gamma > 2\bar{J}$, since in this case $\lim_{s\rightarrow \infty} \beta_s =0$. This is the prerequisite for the bound state and is the key assumption behind the solution (see Appendix C). For $\gamma \leq 2\bar{J}$ there is no bound state.

Let us analyse the properties of the ground state. When $\gamma\gg 2\bar J$ the term $\beta_1$ becomes much larger than any other $\beta_s$ and the ground state tends to a four-partite state describable by $q_{1,0}^{\dagger}|0\rangle$. On the other hand, for $\gamma \rightarrow 2\bar{J}$ all coefficients become equal, i.e., $\beta_s \rightarrow B$. This corresponds to two bipartite composite bosons represented by the state (\ref{2bipartite}) or (\ref{biasfour}). 

It is somehow surprising that the second case occurs for non-zero $\gamma$. In order to understand this effect let us recall the form of (\ref{Heff}) and the relations (\ref{hardcore}). In order to have (\ref{2bipartite}) as a ground state each configuration of two pairs $\eta_k^{\dagger}\eta_{k'}^{\dagger}$ needs to contribute with the same amount of energy. More precisely, we want the ground state to be of the form
\begin{equation}
\mathcal{N}\sum_{k<k'}\eta_k^{\dagger}\eta_{k'}^{\dagger}|0\rangle
\end{equation} 
where $\mathcal{N}$ is the normalization constant. After the application of (\ref{Heff}) the state changes to  
\begin{equation}
\mathcal{N}\sum_{k<k'}\varepsilon_{k,k'}\eta_k^{\dagger}\eta_{k'}^{\dagger}|0\rangle,
\end{equation} 
where $\varepsilon_{k,k'}$ are different energy contributions. However, the above state should be of the same form as before. Only the normalization constant may change. But this requires that $\varepsilon_{k,k'}$ is the same for all $k$ and $k'$. If the interaction term were zero, the kinetic energy would result in $\varepsilon_{k,k'}=-4\bar{J}$ for $k'\neq k+1$ and $\varepsilon_{k,k+1}=-2\bar{J}$. This is due to Pauli exclusion principle, which imposes (\ref{hardcore}). To compensate this, we need to add the nearest neighbour interaction of the strength $-\gamma=-2\bar{J}$.


\subsection{Numerical simulations}

\begin{figure}[t]
\includegraphics[width=0.5 \textwidth,trim={2cm 1cm 1cm 1cm},clip]{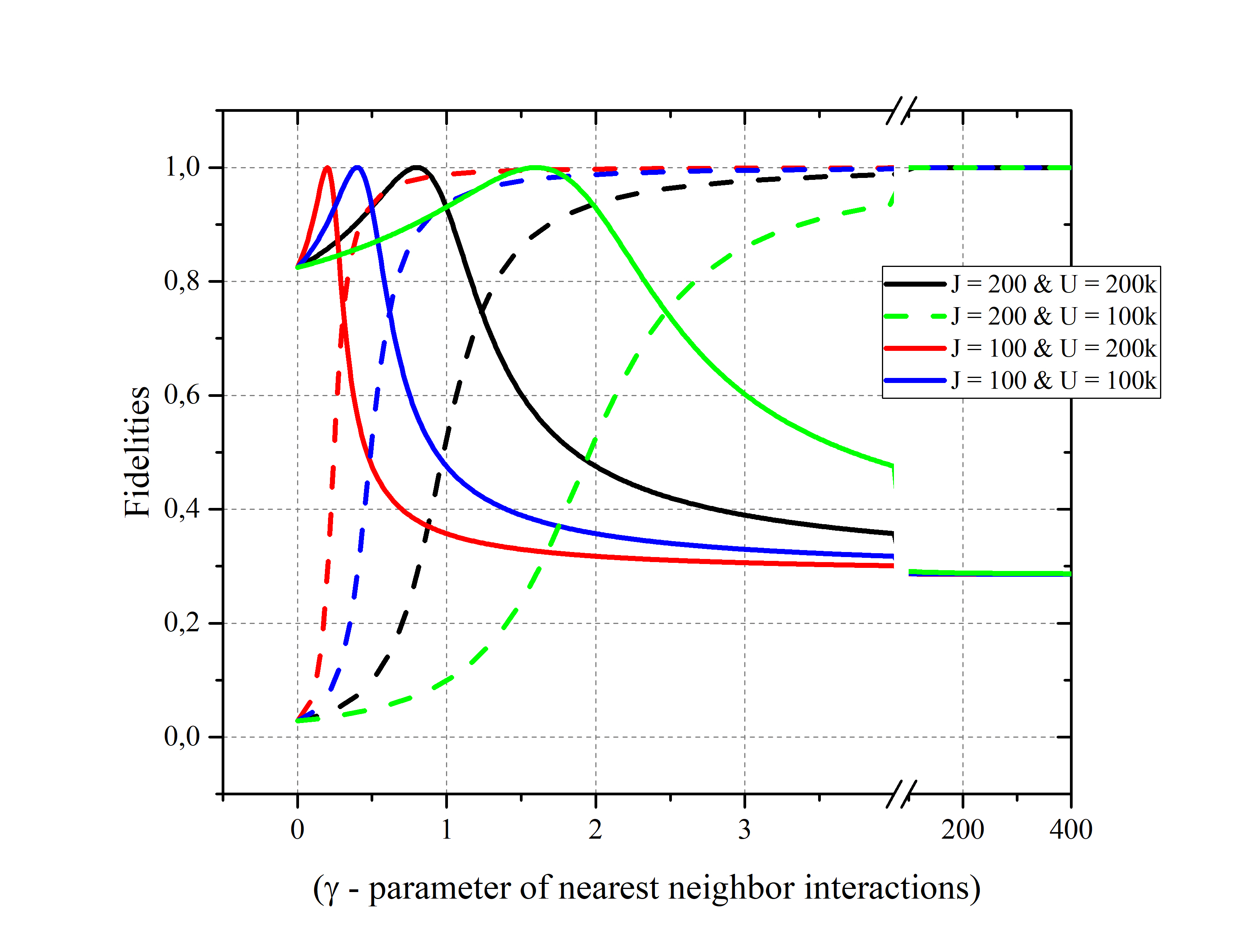}
\caption{Dependence of fidelities $|\langle\psi(\gamma)|q_{1,0}^{\dagger}|0\rangle|^2$ (dashed) and $|\langle\psi(\gamma)|\frac{c^{\dagger 2}_{0,0}}{\sqrt{2\chi_2}}|0\rangle|^2$ (solid) on parameter $\gamma$ for few fixed values of $J$ and $U$. \label{fig2}}
\end{figure}

In order to confirm our predictions, we studied numerically the ground state in the limit $U\gg J\gg\gamma$ and for $d=8$ (assuming periodic boundary conditions). We found that the results of numerical simulations using (\ref{H}) coincide with the ones obtained using (\ref{Heff}). For the fixed $J$ and $U$ we studied how the ground state $|\psi(\gamma)\rangle$ changes as the parameter $\gamma$ increases. We calculated the fidelities $|\langle\psi(\gamma)|q_{1,0}^{\dagger}|0\rangle|^2$ and $|\langle\psi(\gamma)|\frac{c^{\dagger 2}_{0,0}}{\sqrt{2\chi_2}}|0\rangle|^2$. The results are plotted in Fig. \ref{fig2}. The parameter $U$ was chosen to be of the order $10^5$, $J$ of the order $10^2$ and $\gamma$ of the order of $10^0$. Interestingly, if one fidelity approaches one, the other fidelity approaches $1/4$. This is because of the finiteness of space in our simulations. Note that $|\langle 0|q_{1,0}\frac{c^{\dagger 2}_{0,0}}{\sqrt{2\chi_2}}|0\rangle|^2=d \binom{d}{2}^{-1}$ (see Eqs. (\ref{biasfour}) and (\ref{fourpartite})). In the limit $d\rightarrow \infty$ the two states become orthogonal.

\subsection{Bosonic properties of the four-partite entangled state}

Finally, let us investigate if the state $q^{\dagger}_{1,0}|0\rangle$ can be interpreted as a single four-partite bosonic molecule. We do this by analysing the bosonic quality of the operator $q^{\dagger}_{1,0}$ via the same methods as in \cite{Law}. In particular, we analyse the parameter $\chi^{(2)}_N$ defined by
\begin{equation}
\langle 0|q_{1,0}^{N}q^{\dagger N}_{1,0}|0\rangle = \chi^{(2)}_N N!.
\end{equation}

The operator $q^{\dagger}_{1,0}$ generates the bosonic-like ladder structure analogous to (\ref{ladder}). If the ratio $\frac{\chi^{(2)}_{N}}{\chi^{(2)}_{N-1}} \rightarrow 1$ for all $N$, then $q^{\dagger}$ can be considered perfect bosonic. From the results presented in Appendix D we get that 
\begin{equation}
\chi_N^{(2)} = \frac{N!}{d^N}{{d-N}\choose{d-2N}}
\end{equation}
and
\begin{equation}
\frac{\chi_{N+1}^{(2)}}{\chi_N^{(2)}} = 
\left(1-\frac{N+1}{d}\right)\Pi_{i=1}^{N}\left(1-\frac{2}{d+i-2N}\right).
\end{equation}
The above is upper bounded by 1 and lower bounded by
\begin{equation}
\left(1-\frac{N+1}{d}\right)\left(1-\frac{2}{d+1-2N}\right)^N,
\end{equation}
which in the limit $d\gg N$ approaches 1. Therefore, for $d\gg N$ the action of operator $q^{\dagger}_{1,0}$ can be considered as a creation of a single four-partite bosonic particle.


\section{Multipartite composite bosons}

\begin{figure}[t]
\includegraphics[width=0.2 \textwidth]{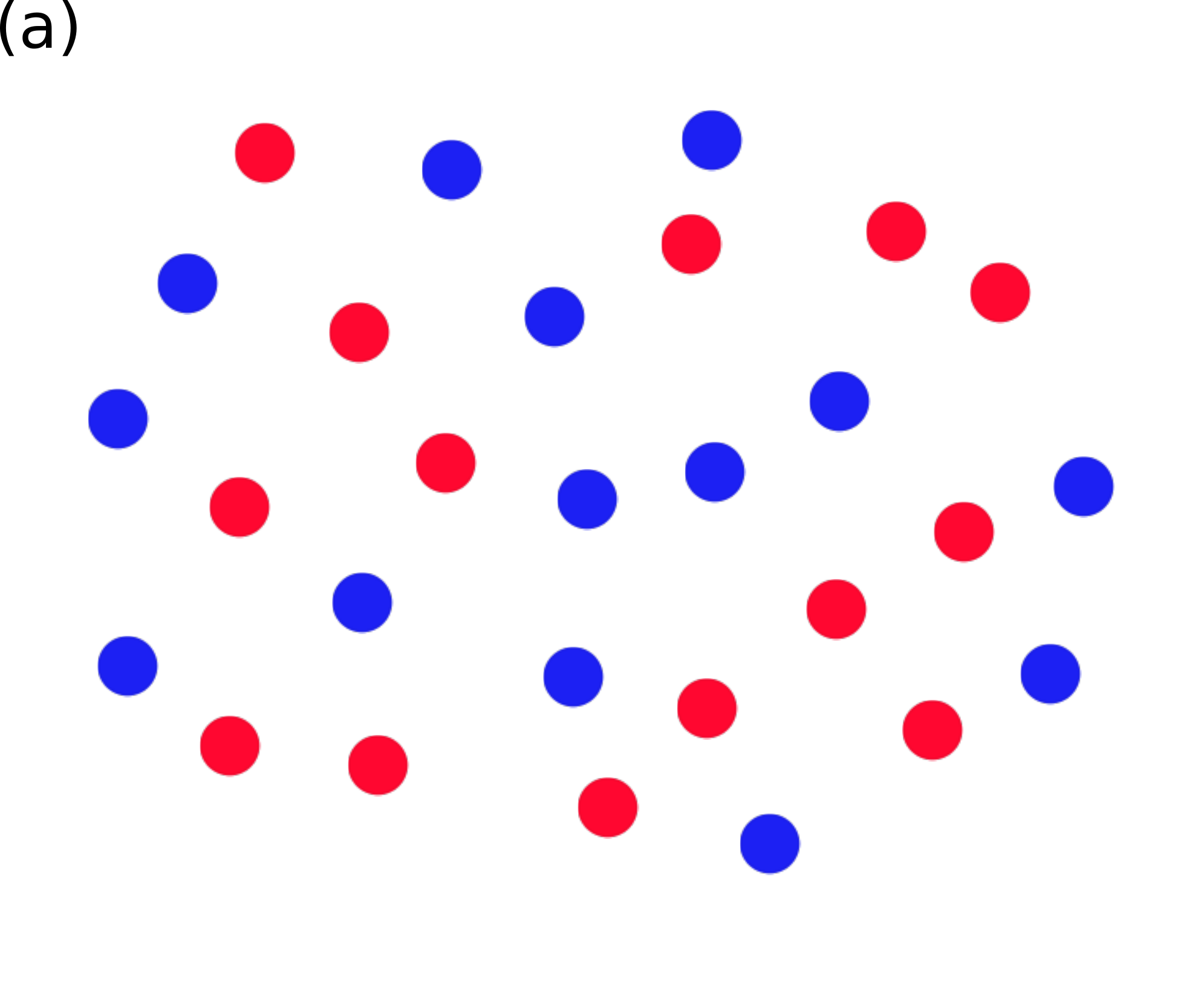}~~\includegraphics[width=0.2 \textwidth]{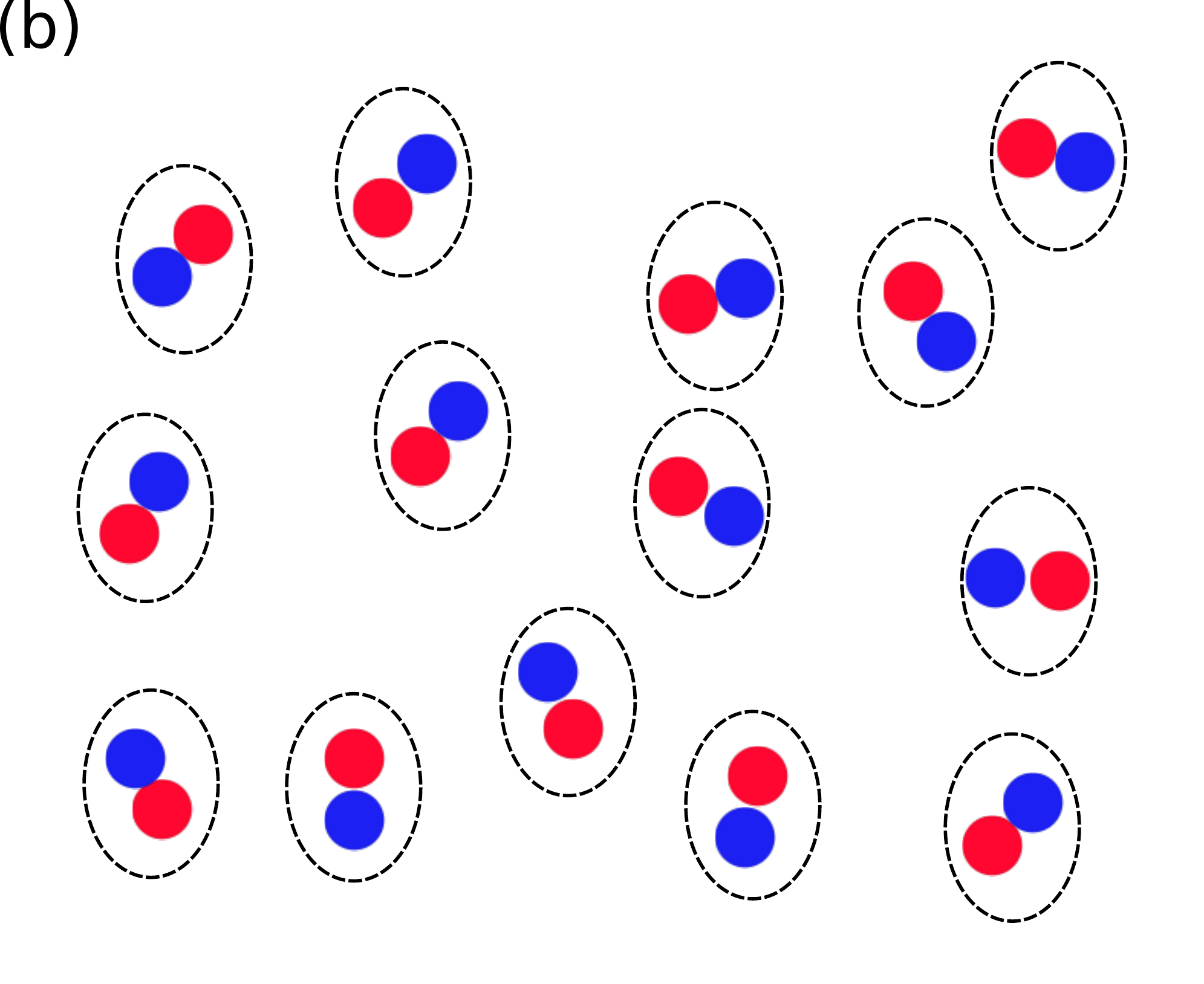}
\includegraphics[width=0.2 \textwidth]{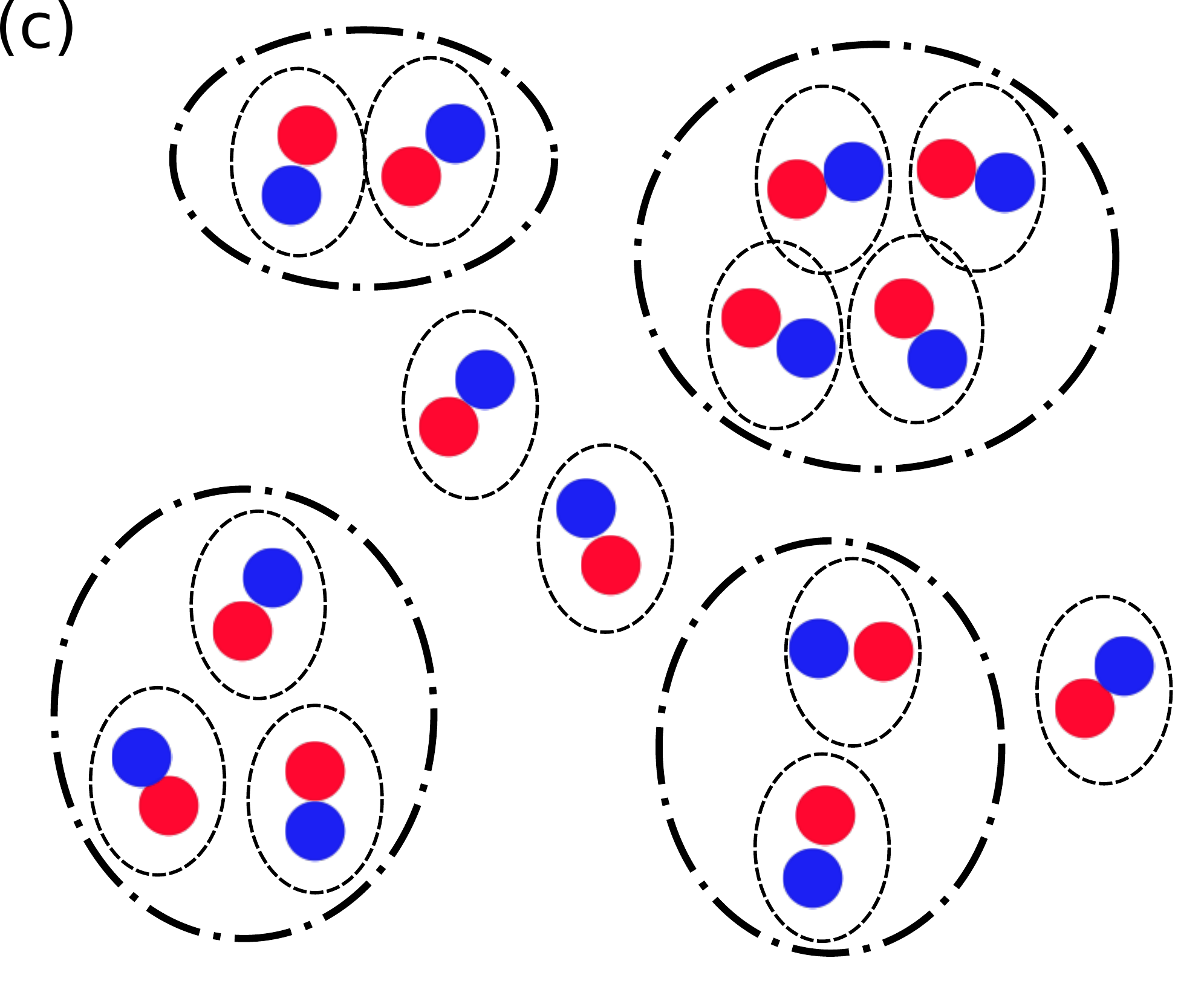}~~\includegraphics[width=0.2 \textwidth]{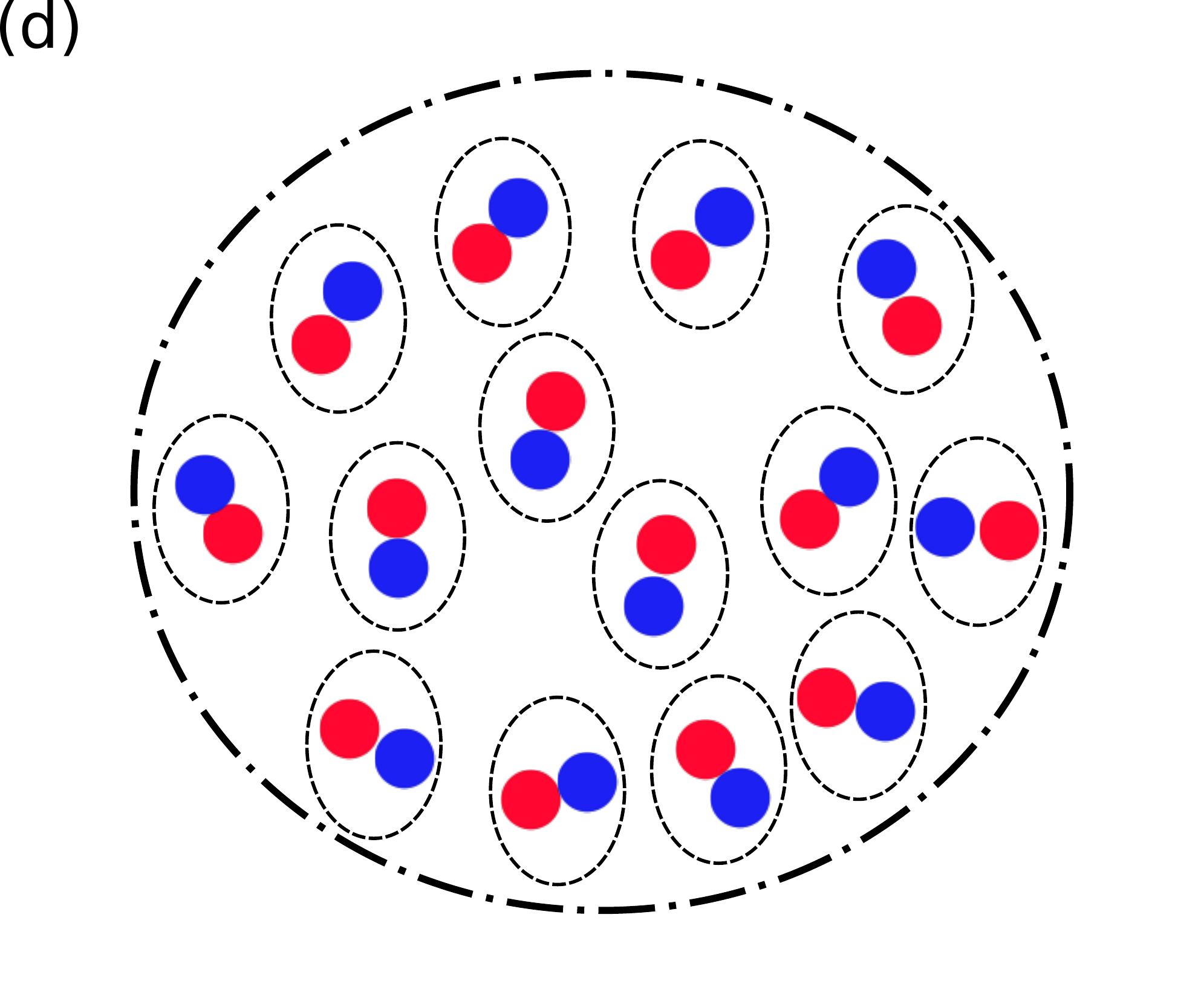}
\caption{Schematic representation of multifermionic assemblies. (a) Free fermions if there are no interactions. (b) In the presence of strong point interactions and weak nearest neighbour interactions bosonic A-B pairs emerge. (c) In the presentce of strong point interaction and nearest neighbour interactions approximatelly equal to the kinetic energy the system starts to assemble into multipartite composite bosons. (d) In the presentce of strong point interaction and strong nearest neighbour interactions a single bosonic molecule is created. \label{fig3}}
\end{figure}

Here we consider composite boson made of $2N$ fermions. Such composite particles are expected to appear in the effective extended Hubbard model (\ref{Heff}) if the particle attraction $\gamma$ is stronger than the effective kinetic energy contribution $\bar J$. On the other hand, if the effective kinetic energy contribution is two times stronger than the attraction between the particles, the system should be describable by $N$ A-B independent pairs. The transition between these two types of behaviour should occur when effective kinetic energy and attraction are of the same strength. A schematic representation of the above multifermionic assemblies are shown in Fig. \ref{fig3}.

\subsection{Multipartite entangled states}

It is natural to expect that for N fermions of type A, N fermions of type B, and sufficiently strong nearest neighbour interaction the Hamiltonian (\ref{Heff}) has the following ground state
\begin{equation}\label{2Mpartite}
q^{\dagger}_{(M)}|0\rangle \equiv \frac{1}{\sqrt{d}}\sum_{k=0}^{d-1}\eta_{k}^{\dagger}\eta_{k+1}^{\dagger}\ldots\eta_{k+M-1}^{\dagger}|0\rangle.
\end{equation}
Note, that for $M=1$ the above state becomes the bi-partite state (\ref{bfbasis}) corresponding to $c^{\dagger}_{0,0}|0\rangle$ and for $M=2$ it becomes the four-partite state (\ref{fourpartite}) corresponding to $q^{\dagger}_{1,0}|0\rangle$. Therefore, the above expectation is in accordance with the few particle cases studied above.


\subsection{Bosonic quality of multipartite entangled states}

The state $q^{\dagger}_{(M)}|0\rangle$ can be considered a single composite boson made of $2M$ fermions if $d \gg M$. We provide a detailed proof in the Appendix D. As a result, the state 
\begin{equation}
\frac{q_{(M)}^{\dagger N}}{\sqrt{\chi_N^{(M)}N!}}|0\rangle
\end{equation}
represents $N$ such composite particles, provided $d\gg NM$, where
\begin{equation}
\chi_N^{(M)} = \frac{\Pi_{i=1}^{N}(d-NM+i)}{d^N}.
\end{equation}

The above bosonic behaviour stems from the fact that state $q_{(M)}^{\dagger}|0\rangle$ is multipartite entangled. In fact, multipartite entanglement is a necessary condition for a bosonic behaviour of 2M fermions. Firstly, note that 2M fermions in a fully separable state, i.e., having Slater rank one \cite{Eckert2002},
\begin{equation}
q_{sep}^{\dagger}|0\rangle\equiv a^{\dagger}_{k_1}\ldots a^{\dagger}_{k_M}b^{\dagger}_{k'_1}\ldots b^{\dagger}_{k'_M}|0\rangle,
\end{equation}
cannot be considered a composite boson. The simplest proof is that for such states there is no two-boson state. We get $c_{sep}^{\dagger 2}|0\rangle = 0$, which is a consequence of Pauli exclusion. Next, note that the same happens if at least one fermion is in a well defined mode, for example
\begin{equation}
 \left(a^{\dagger}_{k_1}\sum_{k_2,\ldots,k'_M}\alpha_{k_2,\ldots,k'_M} a^{\dagger}_{k_2}\ldots a^{\dagger}_{k_M}b^{\dagger}_{k'_1}\ldots b^{\dagger}_{k'_M}\right)^2|0\rangle = 0.
\end{equation} 
Therefore, each fermion needs to be entangled to prevent $q_{(M)}^{\dagger 2}=0$. This implies multipartite entanglement. However, multipartite entanglement is not enough. In Appendix E we show that a multifermionic state representing a single composite boson needs to be genuinely multipartite entangled. This is the only way to recover the ladder structure (\ref{ladder}) of composite bosonic operators. 


\subsection{Composite bosons of various sizes}

Let us define the following states of 2N fermions representing $k$ composite bosons of various sizes
\begin{equation}
|M_1 + \ldots + M_k\rangle \equiv \mathcal{N} q^{\dagger}_{(M_1)}\ldots q^{\dagger}_{(M_k)}|0\rangle,
\end{equation}
where we assume the decreasing order, i.e., $M_1\geq \ldots \geq M_k$. Moreover, $M_1+\ldots+M_k=N$ and $\mathcal{N}$ is a normalization factor. This normalization factor is necessary due to two reasons. Firstly, it may happen that in the above state there are $m$ composite bosons of the same size, i.e., $M_i=\ldots=M_{i+m}=M$. In this case, if there were no normalization factor the norm of the state would be proportional to $\chi_m^{(M)}m!$. Secondly, even if all composite bosons were of a different size, the state $q^{\dagger}_{(M_1)}\ldots q^{\dagger}_{(M_k)}|0\rangle$ would not be normalized, despite the fact that each of the states $q^{\dagger}_{(M_1)}|0\rangle,\ldots,q^{\dagger}_{(M_k)}|0\rangle$ is of norm one. This is because of indistinguishability of fermions, Pauli exclusion and finiteness of $d$. For example, consider a state
\begin{equation}
|3+1\rangle = \mathcal{N}\frac{1}{d}\sum_{k,l=0}^{d-1}\eta_{k}^{\dagger}\eta_{k+1}^{\dagger}\eta_{k+2}^{\dagger}\eta_{l}^{\dagger}|0\rangle.
\end{equation} 
It is easy to find (using the already applied methods) that $\mathcal{N}^2=\frac{d^2}{d^2-4d}$, which in the limit $d\rightarrow \infty$ tends to one.

Our next goal is to investigate to which state $|M_1 + \ldots + M_k\rangle$ the ground state of (\ref{Heff}) corresponds to. For weak interaction the ground state should correspond to $|1+\ldots + 1\rangle$, whereas for strong interaction it should correspond to $|N\rangle$. In addition, we are interested in how the one state changes into the other  as the interaction strength $\gamma$ increases.


\subsection{Numerical simulations for N=3 and N=4}

Here, we discuss numerical results for $N=3$ and $N=4$, i.e., composite bosons made of six and eight fermions. As before, we assume the strong point interaction limit, therefore we consider three and four A-B pairs, respectively, to which we apply the effective Hamiltonian (\ref{Heff}).

\begin{figure}[t]
\includegraphics[width=0.5 \textwidth,trim={2cm 1cm 1cm 1cm},clip]{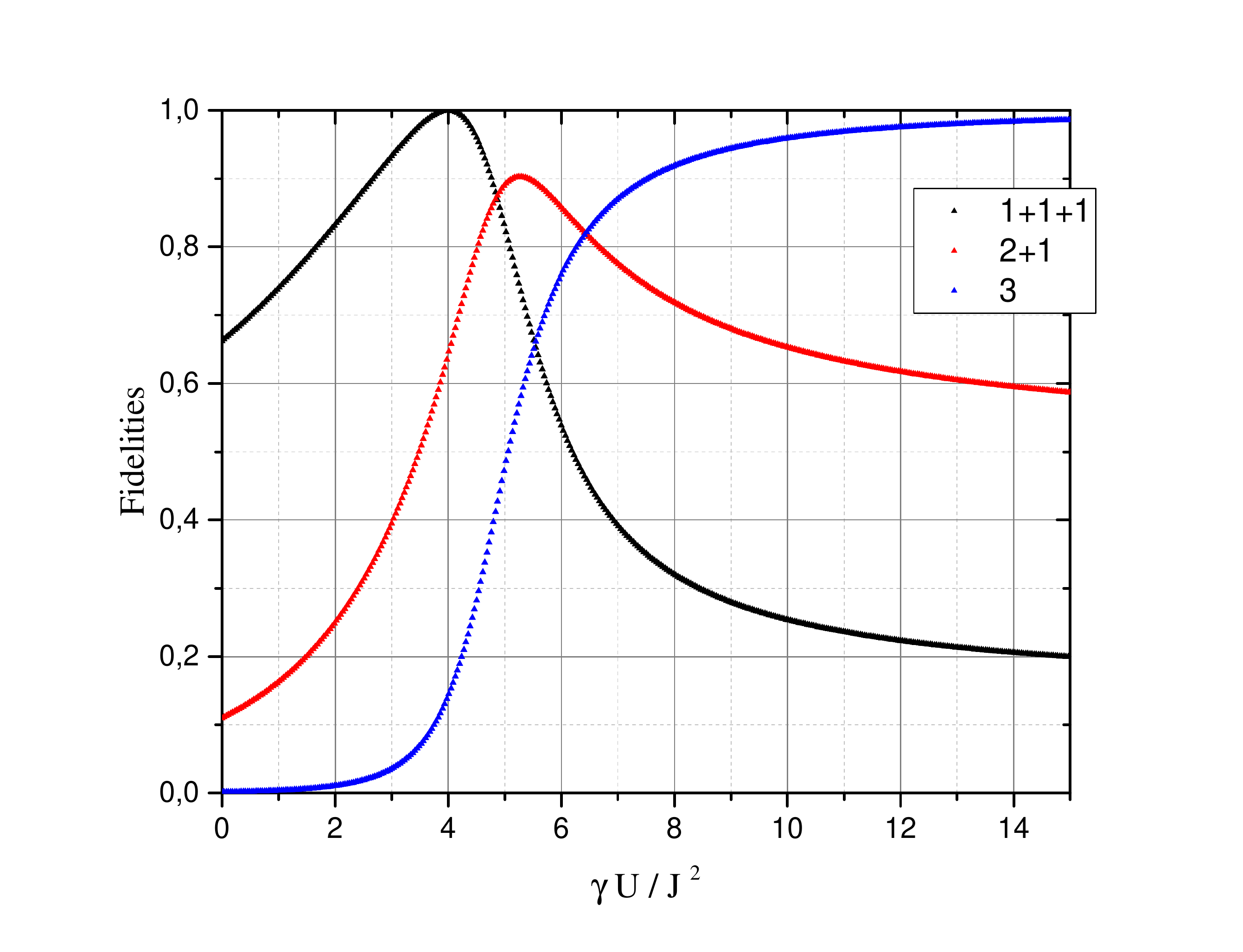}
\includegraphics[width=0.5 \textwidth,trim={2cm 1cm 1cm 1cm},clip]{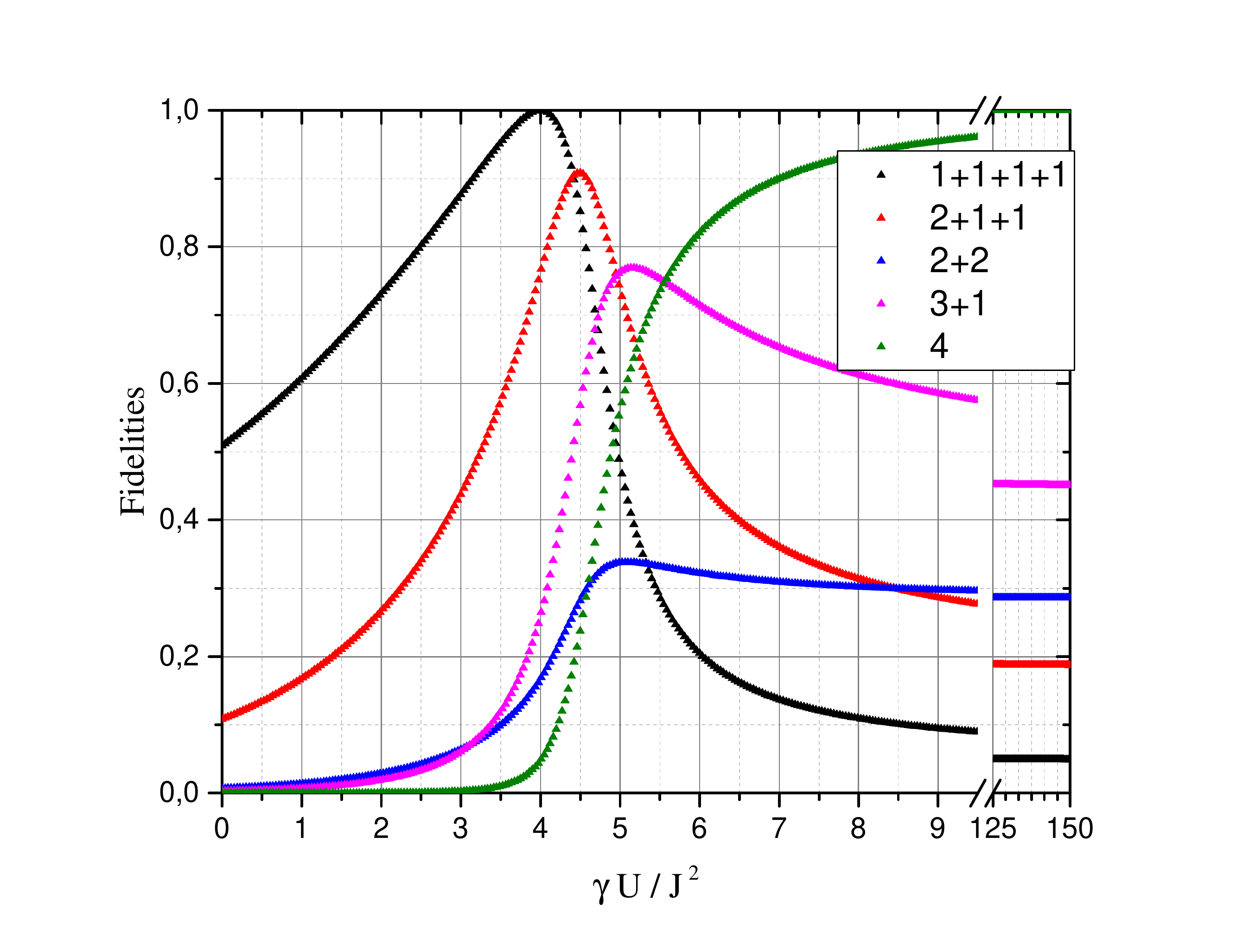}
\caption{Fidelities corresponding to different fermionic assemblies for $N=3$ (top) and $N=4$ (bottom) as functions of $\gamma U/J^2$. \label{fig4}}
\end{figure}

We numerically found the ground state $|\psi(\gamma)\rangle$ and evaluated the fidelities $|\langle \psi(\gamma)|M_1+\ldots+M_k\rangle|^2$. Due to high computational complexity we considered $d=10$, which does not imply a perfect bosonic quality, but still allows us to see some important qualitative behaviour of the model. The corresponding fidelities are plotted in Fig. \ref{fig4}.

In case of $N=3$ we observe that for low value of the ratio $\gamma U/J^2$ the ground state is dominated by the state $|1+1+1\rangle$. The fidelity $|\langle \psi(\gamma)|1+1+1\rangle|^2$ reaches one for $\gamma U/J^2 = 4$. This maximum occurs for $\gamma \neq 0$ due to the same reason as in the case $N=2$. For $\gamma U/J^2 > 4$ there is a small region in which the ground state is dominated by $|2+1\rangle$, although the fidelity of this state never reaches one. After this region the ground state is dominated by $|3\rangle$ and as $\gamma U/J^2 \gg 4$ the fidelity $|\langle \psi(\gamma)|3\rangle|^2 \rightarrow 1$.

In case of $N=4$ the transition from $|1+1+1+1\rangle$ to $|4\rangle$ seems to be more complex. As before, the ground state is dominated by $|1+1+1+1\rangle$ for low values of $\gamma U/J^2$ and the corresponding fidelity reaches maximum of one for $\gamma U/J^2 = 4$. As the ratio increases, the state $|2+1+1\rangle$ and then $|3+1\rangle$ starts to take over, but soon the state $|4\rangle$ becomes to dominate and its fidelity approaches one as the ratio becomes large. Interestingly, the state $|2+2\rangle$ seems to play no significant role in the above transition. The fact that it appears in the Fig. \ref{fig4} is rather due to its overlap with other states, which is caused by the relatively small value of $d$. Note, that for high values of $\gamma U/J^2$ the ground state is describable by $|4\rangle$, but the fidelities corresponding to other states are still high. In particular, the fidelity $|\langle \psi(\gamma)|2+2\rangle|^2$ does not change much after it reaches its maximum of $\approx 0.3$. 

The above observations allow us to speculate that the transition from N A-B pairs to a single N-partite bosonic molecule follows the pattern,
\begin{equation}\label{transition}
|1+1+\ldots\rangle \rightarrow |2+1+\ldots\rangle \rightarrow |3+1+\ldots\rangle \rightarrow \ldots \rightarrow |N\rangle.
\end{equation}  
This means that bi-partite bosonic particles assemble into a large bosonic molecule by adding particle by particle into a single large compound. In this situation one does not observe creation of two, or more, larger compounds. In the next subsection we support this hypothesis by showing that the transition (\ref{transition}) is most energetically favourable. It is worth to add here that during \textit{transition} process (when strength of $\gamma$ is increased) we assume that system follows its ground state.


\subsection{Transition from N bipartite composite bosons to a single bosonic molecule}

Before we go into details, let us first discuss an important property of states $|M_1+\ldots+M_k\rangle$. They are superpositions of terms 
\begin{equation}
\eta^{\dagger}_{j_1}\ldots \eta^{\dagger}_{j_1+M_1-1}\ldots\eta^{\dagger}_{j_k}\ldots \eta^{\dagger}_{j_k+M_k-1}|0\rangle.
\end{equation}
The sequence $\eta^{\dagger}_{j_i}\ldots \eta^{\dagger}_{j_i+M_i-1}$ represents the i'th compound made of $M_i$ A-B pairs. There is a possibility that in this superposition some two compounds are next to each other, i.e., for  $\eta^{\dagger}_{j_i}\ldots \eta^{\dagger}_{j_i+M_i-1}$ and $\eta^{\dagger}_{j_{i+1}}\ldots \eta^{\dagger}_{j_{i+1}+M_{i+1}-1}$ we have $j_{i+1}=j_i+M_i$. Nevertheless, the total number of terms in the superposition is of the order $d^k$, whereas the number of terms with adjacent compounds scales as $d^{k-1}$. Therefore, for $d\gg N$ one can assume that almost all terms in the superposition correspond to non-adjacent compounds (non-adjacency assumption). This leads to
\begin{equation}
\langle M'_1+\ldots+M'_{k'}|M_1+\ldots+M_k \rangle = 0
\end{equation}  
for two different configurations $\{M_1,\ldots,M_k\}$ and $\{M'_1,\ldots,M'_{k'}\}$, although the total number of particles in both configurations is the same.

Let us once more consider the Hamiltonian (\ref{Heff}) and estimate its expectation value for a state $|M_1+\ldots+M_k\rangle$
\begin{equation}
\langle \mathcal{H}_{eff} \rangle = \langle \mathcal{H}_{k} \rangle + \langle  \mathcal{H}_{p} \rangle,
\end{equation}
where we explicitly split the kinetic and the potential energy parts. First, we consider the kinetic part. Note that under the action of $\mathcal{H}_{k}$ the compounds made of more than a single A-B pair split into smaller compounds, i.e.,
\begin{eqnarray}
& &\mathcal{H}_{k}\eta^{\dagger}_{j_i}\eta^{\dagger}_{j_i+1}\ldots \eta^{\dagger}_{j_i+M_i-2}\eta^{\dagger}_{j_i+M_i-1}|0\rangle = \\
& &-\bar{J}\eta^{\dagger}_{j_i-1}\eta^{\dagger}_{j_i+1}\ldots \eta^{\dagger}_{j_i+M_i-2}\eta^{\dagger}_{j_i+M_i-1}|0\rangle \nonumber \\
& &-\bar{J}\eta^{\dagger}_{j_i}\eta^{\dagger}_{j_i+1}\ldots \eta^{\dagger}_{j_i+M_i-2}\eta^{\dagger}_{j_i+M_i}|0\rangle, \nonumber
\end{eqnarray} 
where $\bar{J}=2J^2/U$. Therefore, under the action of $\mathcal{H}_{k}$ the state $|M_1+\ldots+M_k\rangle$ changes into superposition of states $|M'_1+\ldots+M'_{k'}\rangle$. Interestingly, if in the original state there were no single A-B pairs ($M_k > 1$), then none of the states $|M'_1+\ldots+M'_{k'}\rangle$ in the effective superposition is equal to the original one. Therefore, the non-adjacency assumption implies
\begin{equation}
\langle M_1+\ldots+M_k|\mathcal{H}_{k}|M_1+\ldots+M_k \rangle = 0~~~\text{if}~~~M_k>1.
\end{equation}

Next, assume that in $|M_1+\ldots+M_k\rangle$ there are $r$ single A-B pairs. The kinetic energy term moves these pairs one step to the right and one step to the left. Recall that the above state consists of a superposition of all possible (non-adjacent) configurations of such pairs. Therefore, 
\begin{eqnarray}
& &\langle M_1+\ldots+M_k|\mathcal{H}_{k}|M_1+\ldots+M_k \rangle = -2r\bar{J} \nonumber \\
& &\text{if}~~~~M_{k-r+1}=\ldots=M_k=1. 
\end{eqnarray}
This is because each term in $|M_1+\ldots+M_k\rangle$ can be obtained from $2r$ other therms by shifting some A-B pair either one step to the right or one step to the left.

The expectation value of the potential energy part is much easier to evaluate. It is straightforward to show that under the non-adjacency assumption
\begin{equation}
\langle M_1+\ldots+M_k|\mathcal{H}_{p}|M_1+\ldots+M_k \rangle = -(N-k)\bar{\gamma},
\end{equation}
where $\bar{\gamma}=2(\gamma-\bar{J})$. Interestingly, this value depends only on the total number of compounds $k$, not on the way the A-B pairs are distributed between these compounds $\{M_1,\ldots,M_k\}$. As a result
\begin{equation}
\langle \mathcal{H}_{eff} \rangle = -2r\bar{J} -(N-k)\bar{\gamma}.
\end{equation}

Now, consider two states with the same number of compounds $k$, but different number of single A-B pairs, $r$ and $r'<r$ (e.g. $|3+1+1\rangle$ and $|2+2+1\rangle$). The corresponding average energies are $-2r\bar{J} -(N-k)\bar{\gamma}$ and $-2r'\bar{J} -(N-k)\bar{\gamma}$. It is clear that the average energy is lower for the state containing more single A-B pairs. Therefore, only the states of the form $|M+1+\ldots+1\rangle$ need to be taken into account during the transition between $|1+\ldots+1\rangle$ into $|N\rangle$, which confirms our previous hypothesis. For such states the average energy equals
\begin{eqnarray}
& &\langle M+1\ldots+1|\mathcal{H}_{eff}|M+1\ldots+1\rangle = \nonumber \\ 
& &-(M-1)\bar{\gamma}-2(N-M+\delta_{M,1})\bar{J}, \label{energy}
\end{eqnarray}  
where $\delta_{M,1}$ is the Kronecker delta. 

\begin{figure}[t]
\includegraphics[width=0.45 \textwidth,trim={0cm 0cm 0cm 0cm},clip]{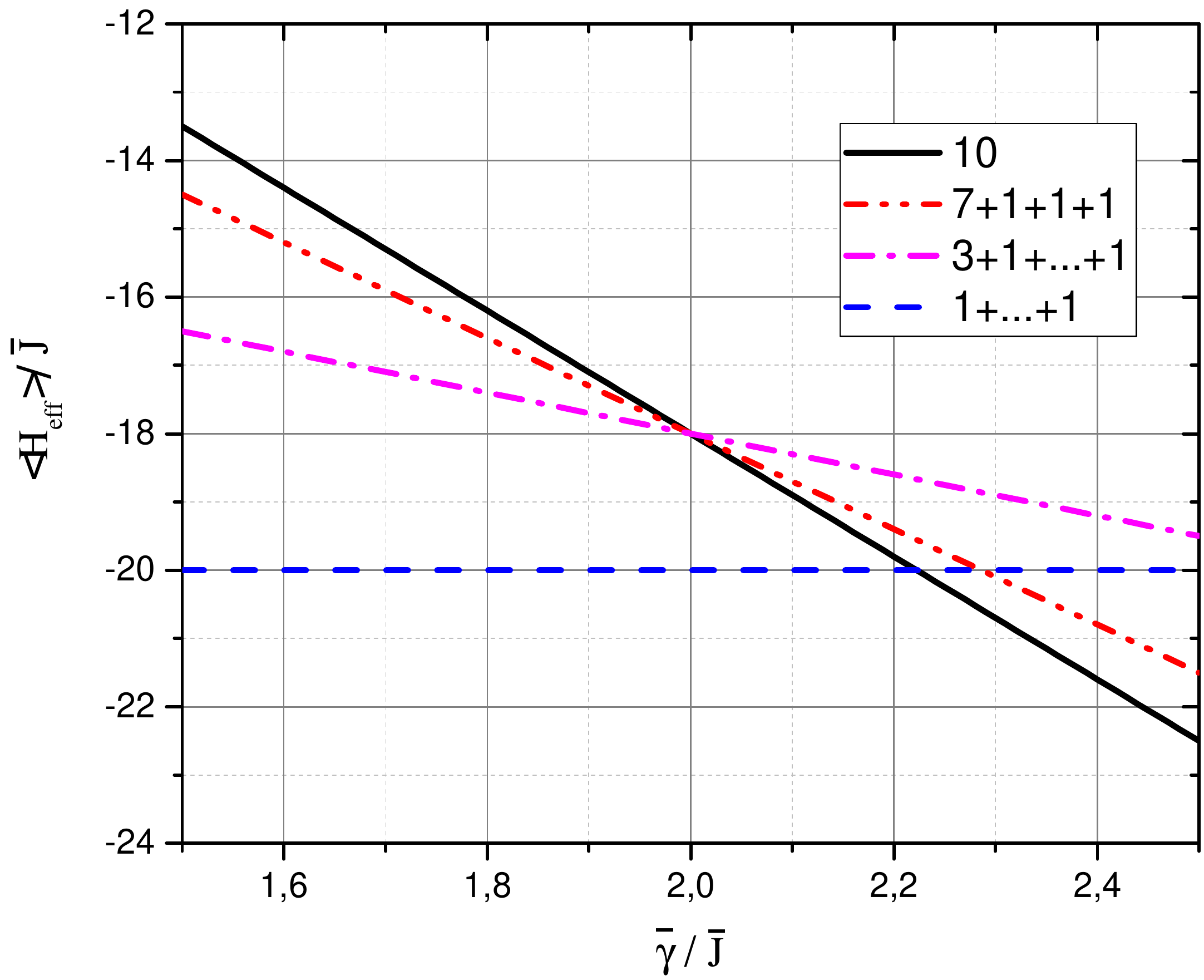}
\caption{Plot of the average energy for $N=10$ and states $|10\rangle$, $|7+1+\dots+1\rangle$, $|3+1+\dots+1\rangle$, and $|1+1+\ldots+1\rangle$. In the limit $d\gg N$ for which the no-adjacency assumption is valid the lowest energy corresponds either to $|1+1+\ldots+1\rangle$ or to $|10\rangle$, which suggests that in this case one state is directly transformed into the other as the strength of the nearest neighbour interaction increases. \label{fig5}}
\end{figure}

In Fig. \ref{fig5} we plot few of these values for $N=10$. The lowest energy corresponds either to $|1+1+\ldots+1\rangle$ or to $|10\rangle$. This observation implies the following conjecture: for $d\gg N$ (no-adjacency assumption) the state $|1+1+\ldots+1\rangle$ is directly transformed into $|N\rangle$ as $\gamma$ increases. Referring to Fig. \ref{fig3}, the above means that the assembly (b) goes directly to (d) and the assembly (c) never occurs. Using the formula (\ref{energy}) one can show that the transition should occur for $\bar{\gamma}/\bar{J}=2N/(N-1)$, which corresponds to $\gamma U/J^2 = 2+ 2N/(N-1)$. This agrees with plots in Fig. \ref{fig4}. From the point of view of entanglement analysis, the conjecture implies that the system of 2N bipartite entangled fermions is directly transformed into genuinely multipartite entangled state.


\subsection{Entanglement and correlation function between fermion pairs}

\begin{figure}[h]
\includegraphics[width=0.43 \textwidth]{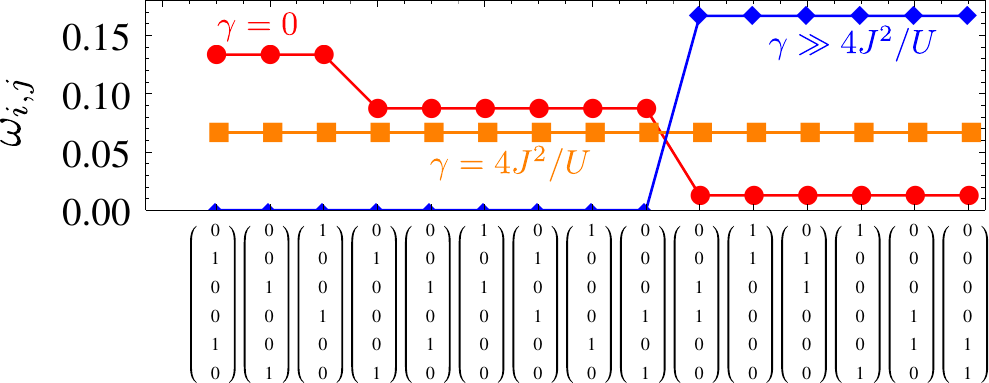}
\caption{Ground state $|\psi(\gamma)\rangle = \sum_{i<j} \sqrt{\omega_{i,j}} \eta_i^\dagger \eta_j^\dagger|0\rangle$ of $\mathcal{H}_\text{eff}$, Eq. \eqref{Heff}, for $N=2$ and $d=6$. Red dot are the weights $\omega_{i,j}$ for $\gamma = 0$, orange squares for $\gamma =4J^2/U$ and blue diamonds for  $\gamma \gg4J^2/U$.}
\label{FigStateEBH}
\end{figure}

In general, in the ground state $|\psi(\gamma)\rangle$ the $N$ A-B pairs are correlated. For $\gamma = 0$ an effective repulsive interaction between them, inherent in one-dimensional fermion systems, leads to particle anti-bunching (see red dot in Fig. \ref{FigStateEBH}). This repulsive interaction is compensated by a nearest neighbour interaction of strength $\gamma =4J^2/U$ (orange squares) -- in this case the pairs independently occupy the sites of the lattice. For large  interaction strength $\gamma \gg4J^2/U$ (blue diamonds) the pairs become maximally multipartite entangled.

Entanglement has been demonstrated to be useful for the identification of quantum phase transitions in the extended 1D Hubbard model \cite{ShiJian2004}. Here we will show that the entanglement between fermion pairs can be used to identify the transition between bunching and anti-bunching. 

\begin{figure}[t]
\includegraphics[width=0.45 \textwidth]{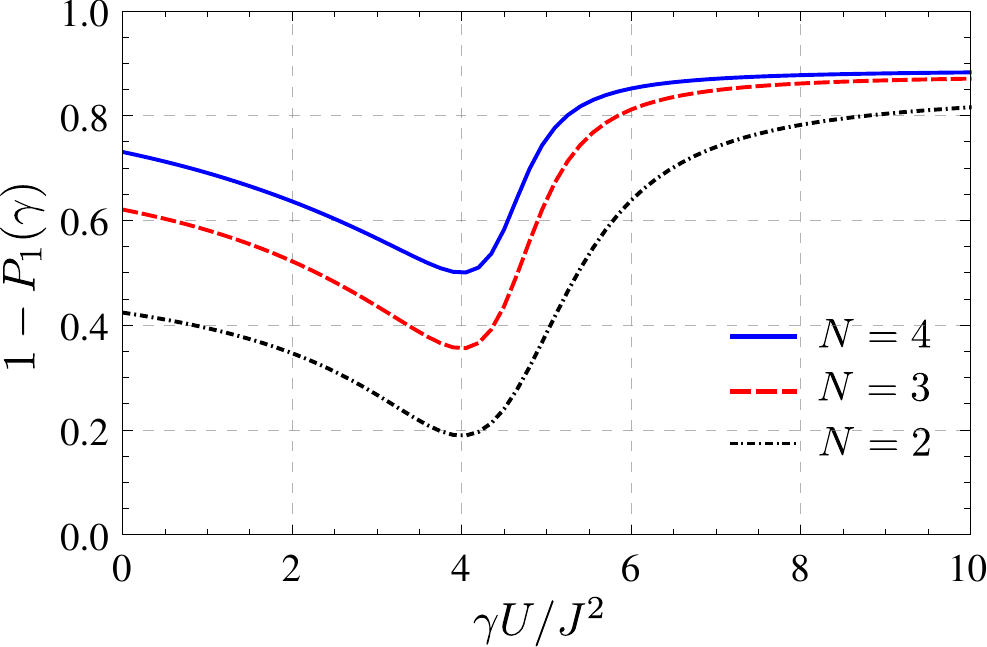}
\caption{Entanglement $1-P_1(\gamma)$ as a function of $\gamma U/J^2$ for a lattice of $d=10$ sites and $N=2,3,4$, blue, red dashed and black dotted dashed lines respectively}
\label{FigEnt}
\end{figure}

Since fermion pairs are identical, the amount of entanglement can be characterized by the purity 
\begin{equation}
P_1(\gamma)=\text{Tr}\left[\left(\rho_{ij}^{(1)}\right)^2\right] = \sum_{ij=0}^{d-1}\rho_{ij}^{(1)}\rho_{ji}^{(1)}
\end{equation}
of the single-pair reduced density matrix 
\begin{equation}
\rho_{ij}^{(1)}= \frac{1}{N} \langle\psi(\gamma)|\eta_i^\dagger \eta_j|\psi(\gamma)\rangle.
\end{equation}
Note that, as in case of indistinguishable fermions \cite{BouvrieValdesetall2016}, the normalization factor of the reduced density matrix $\rho_{ij}^{(1)}$ is $1/N$. We show in Fig. \ref{FigEnt} the entanglement $1-P_1(\gamma)$ between one pair and the remaining $N-1$ pairs. The correlations are minimized for the interaction $\gamma U/J^2=4$. In this interaction regime the purity has the analytical solution 
\begin{equation}
P_1(4J^2/U) = \frac{1}{d}+\frac{(d-N)^2}{d(d-1)}.
\end{equation}
In the limit $d\to\infty$ the ground state corresponds to $N$ independent pairs ($P_1(4J^2/U)=1$), where each pair is maximally entangled. On the other hand, the maximal correlations are reached for $\gamma \gg 4J^2/U$, for which 
\begin{equation}
P_1(\infty) = \frac{1}{d} \times \left\{ \begin{array}{cc}
               1+2/N^2 & \text{~~ if ~~} d>2N \\
               1+4/N^2 & \text{~~ if ~~} d= 2N \\
               1+2/N^2 & \text{~~ if ~~} 2N>d > N+2 \\
               1 & \text{~~ if ~~}  N + 2 \ge d 
              \end{array}
 . \right. 
\end{equation}
The entanglement takes its maximum value in the limit $d\to \infty$, where $P_1(\infty) =0$. In this case all particles are maximally multipartite entangled.

\begin{figure}[h]
\includegraphics[width=0.45 \textwidth]{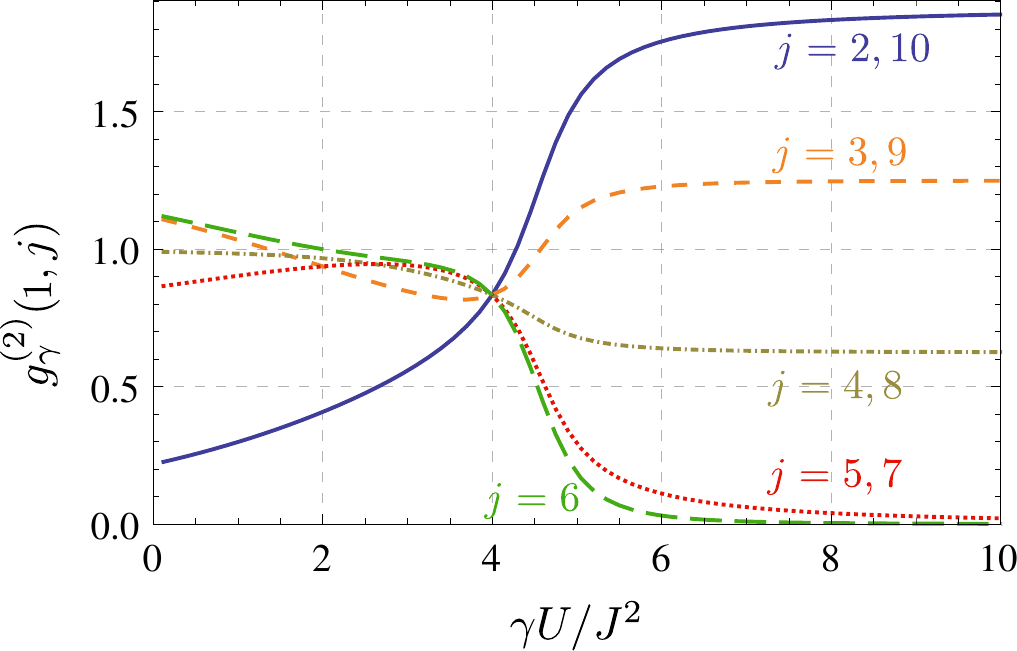}
\caption{Second order correlation function $g^{(2)}_\gamma(i,j)$ as a function of $\gamma U/J^2$ for $N=4$ and $d=10$.}
\label{FigCorrFunc}
\end{figure}

The above has observable consequences in the second order correlation function
\begin{equation}
g^{(2)}_\gamma(i,j) = \frac{\langle\psi(\gamma)|\eta_i^\dagger \eta_j^\dagger \eta_i \eta_j|\psi(\gamma)\rangle}{\langle\psi(\gamma)|\eta_i^\dagger \eta_i |\psi(\gamma)\rangle^2}
\end{equation}
of the fermions pairs located at positions $i$ and $j$. The mean occupations of sites are constant $\langle\psi(\gamma)|\eta_i^\dagger \eta_i |\psi(\gamma)\rangle=N/d$. As can be seen in Fig. \ref{FigCorrFunc}, small nearest neighbour interaction yields anti-bunching between nearest neighbours $g^{(2)}_\gamma(i,i+1)< 1$. For $\gamma = 4J^2/U$ the anti-bunching is only due to the Pauli exclusion principle between identical pairs $g^{(2)}_{4J^2/U}(i,j) = \frac{d (N-1)}{N(d-1)} ~~~\forall i,j$ and in the limit $d\to\infty$ the usual boson correlation function $g_\text{bosons}^{(2)} = (N-1)/N$ is observed. Finally, in the limit of large interaction $g^{(2)}_\infty(i,j) = d(N-|i-j|)/N$ if $|i-j|\le N$ and $g^{(2)}_\infty(i,j) =0$ otherwise. Therefore, bunching dominates the particle statistics, even for long range particle correlations $g^{(2)}_\infty(i,j)>1$ for $(|i-j|)<N-N/d$. Comparing figures \ref{FigEnt} and \ref{FigCorrFunc}, the minimal correlations clearly indicate the transition between bunching and anti-bunching.

\section{Summary}

We studied multifermionic states and showed that in order to form a composite boson made of 2N fermions the system needs to exhibit a genuine 2N-multipartite fermionic entanglement. Next, we tested our results in a specific physical model. We examined emergence of multipartite composite bosons in the ground state of the extended one-dimensional Hubbard model. In particular, we focussed on which different bosonic assemblies emerge as the intra-particle interaction increases. We solved the problem analytically for two and four fermions. In the second case we assumed the strong point interaction limit $U\gg J$. For higher number of fermions we also considered the strong point interaction limit and performed numerical simulations. They confirmed that if the nearest neighbour interactions are strong the system forms a single multipartite bosonic molecule, whereas if they are weak the system consists of many bipartite bosonic particles. Then, we made an analytical estimation of the average energy for different fermionic assemblies. This lead us to the following conjecture: as the nearest neighbour interaction increases it is energetically favourable that the fermions transform from the state representing many bipartite composite bosons into the state corresponding to a single multipartite composite boson, without going through other possible assemblies. This conjecture should be directly verifiable using the analytical methods of Lieb and Wu based on Bethe ansatz \cite{Lieb1968}. 

{\it Acknowledgements.} We would like to thank Ana Majtey, Cecilia Cormick and Paula Cespedes for sharing relevant information concerning the effective Hamiltonian (Eq.~\eqref{Heff}), Marcin Karczewski and Andrzej Grudka for stimulating discussions and suggestions. Z.K., A.S. and P.K. were supported by the National Science Center in Poland through NCN Grant No. 2014/14/E/ST2/00585. P.A.B. gratefully acknowledge support by the Conselho Nacional de Desenvolvimento Cient\'ifico e Tecnol\'ogico do Brasil and by the Spanish MINECO project FIS2014-59311-P (co-financed by FEDER).


\section*{Appendix A}

Here we provide solution to the set of recurrence equations (\ref{as}) and (\ref{a0}). The typical equations (\ref{as}) have general solution of the form
\begin{equation}
\alpha_s = Ar_0^{s} + Br_0^{-s},
\end{equation}
where $r_0$ and $r_0^{-1}$ are roots of the following quadratic equation
\begin{equation}
r^2+\frac{\epsilon}{2J} r + 1 = 0,
\end{equation}
and $A$ and $B$ are constants. In addition
\begin{equation}\label{epsilon}
\epsilon=-2J(r_0+r_0^{-1}).
\end{equation} 
Note, that the ground state should be a bound state. Such states are normalizable, therefore we have the commonly used constraint $\lim_{s \rightarrow \pm \infty}\alpha_s = 0$. Since $r_0^{-1}$ is the inverse of $r_0$, the modulus of one of them is less than one, whereas the modulus of the other one is greater than one. Say, $|r_0|\leq 1$. Therefore, due to the normalization constraint we get
\begin{eqnarray}
\alpha_s &=& Ar_0^{s}~~~~~~~(s>0), \\
\alpha_s &=& Br_0^{-s}~~~~~(s<0).
\end{eqnarray}
To determine $\alpha_0$ we plug the above and (\ref{epsilon}) to 
\begin{eqnarray}
-\frac{\epsilon}{2J}\alpha_1 &=& \alpha_{2} + \alpha_{0}, \\
-\frac{\epsilon}{2J}\alpha_{-1} &=& \alpha_{0} + \alpha_{-2},
\end{eqnarray}
and obtain
\begin{eqnarray}
A(r_0^2 + 1) &=& Ar_0^2 + \alpha_0, \\
B(r_0^2 + 1) &=& Br_0^2 + \alpha_0,
\end{eqnarray}
which imply $A=B=\alpha_0$. 

Finally, we plug the above results to the atypical equation (\ref{a0}) and obtain
\begin{equation}
-\bar{U}+r_0+r_0^{-1}=2r_0,
\end{equation}
where $\bar{U}=U/2J$, which leads to the following quadratic equation for $r_0$
\begin{equation}
r_0^2+\bar{U}r_0-1=0.
\end{equation}
We get
\begin{equation}
r_0=-\frac{1}{2}(\bar{U}\pm\sqrt{\bar{U}^2+4}).
\end{equation}
Since we assumed $|r_0|\leq 1$ we choose the solution with minus sign and get the final form of the ground state
\begin{equation}
\alpha_s = A\left(\frac{\sqrt{U^2+16J}-U}{4J}\right)^{|s|},
\end{equation}
and the corresponding energy
\begin{equation}
\epsilon = -\sqrt{U^2+16J^2}.
\end{equation}


\section*{Appendix B}

Here we derive the effective Hamiltonian (\ref{Heff}). We start with the original one (\ref{H}), which is of the form 
\begin{equation}
\mathcal{H} =  \mathcal{H}_0 +  \mathcal{H}_p +  \mathcal{H}_{nn}
\end{equation}

First, lets consider the Hamiltonian $\mathcal{H}^{(1)} = J \mathcal{H}_0 + U \mathcal{H}_p $. The projector on the ground state of $\mathcal{H}_p$ reads 
\begin{equation}
P_g = \sum_{j_1<j_2<\cdots<j_N} \left( \bigotimes_{n=1}^N \ket{j_nj_n}\bra{j_nj_n} \right),
\end{equation}
where $\ket{j_nj_n}=a_n^{\dagger}b_n^{\dagger}|0\rangle$, and the unitary evolution operator associated to $\mathcal{H}_p$ 
\begin{eqnarray}
 U_g &=& e^{-it \mathcal{H}_p/\hbar}= P_g e^{-it E_g /\hbar} + P_e e^{-it E_e/\hbar}  \nonumber \\ &=& P_g e^{-it N U /\hbar} + P_e e^{-it (N-1) U /\hbar},
\end{eqnarray}
where $P_e$ is the projector on the first exited state which consists of $N-1$ bounded pairs with energy $(N-1)U$ and one separated pair with zero energy. In the Master Thesis of P. Cespedes \cite{cespedes} it has been shown that at first order of perturbation 
\begin{eqnarray}
P_g U_I^{(1)}P_g &=& -\frac{i}{\hbar} \int_0^t P_g H_I(t_1) P_g dt_1  \\
&=& -\frac{i}{\hbar} \int_0^t P_g  U_g(t_1)^\dagger \mathcal{H}_0   U_g(t_1) P_g dt_1 = 0 \nonumber
\end{eqnarray}
since $P_g  H_0 P_g = 0$, and that at second order 
\begin{eqnarray}
&P_g&  U_I^{(2)}P_g = -\frac{i}{\hbar} \int_0^t P_g H_I(t_1)  dt_1 \int_0^{t_1}  H_I(t_2) P_g dt_2  \nonumber \\
&=& -\frac{i}{\hbar} \int_0^t \int_0^{t_1} dt_1 dt_2 P_g \left(  U_g(t_1)^\dagger \mathcal{H}_0   U_g(t_1) \right) \nonumber \\
&\times & \left( U_g(t_2)^\dagger \mathcal{H}_0   U_g(t_2) \right) P_g \nonumber \\
&=& -\frac{i}{\hbar} \int_0^t \int_0^{t_1} dt_1 dt_2 P_g \mathcal{H}_0^2 P_g e^{i(t_2-t_1)U/\hbar} \nonumber \\
&=& -\left(\frac{i}{\hbar}\right)^2 \frac{\hbar}{iU} \left( t+ \frac{\hbar}{iU} e^{-itU/\hbar} -1 \right) P_g \mathcal{H}_0^2 P_g \nonumber \\
& \overset{U\gg J}{\approx}& \frac{i}{U \hbar} P_g \mathcal{H}_0^2 P_g 
\end{eqnarray}
where 
\begin{eqnarray}
P_g \mathcal{H}_0^2 P_g &=& - N 4 J^2 \sum_{k=0}^{d-1}  \eta_k^\dagger  \eta_k - 2 J^2 \sum_{k=0}^{d-1} ( \eta_k^\dagger  \eta_{k+1} + h.c.) \nonumber \\
&+& 4 J^2 \sum_{k=0}^{d-1}  \eta_k^\dagger \eta_k \eta_{k+1}^\dagger \eta_{k+1}.
\end{eqnarray}
The effective Hamiltonian $\mathcal{H}^{(1)}$ for $U \gg J$ is given by
\begin{eqnarray}
\mathcal{H}^{(1)}_\text{eff} &=& - N \left( U + \frac{4 J^2}{U} \right) \sum_{k=0}^{d-1} \eta_k^\dagger \eta_k \nonumber \\
&-& \frac{2 J^2}{U} \sum_{k=0}^{d-1} ( \eta_k^\dagger  \eta_{k+1} + h.c.) \nonumber \\
&+& \frac{4 J^2}{U} \sum_{k=0}^{d-1} \eta_k^\dagger \eta_k \eta_{k+1}^\dagger \eta_{k+1}
\end{eqnarray}
where we have used that $P_g  H_p P_g = - N U$. 

Now let us consider the full Hamiltonian $\mathcal{H}$. At first order we have 
\begin{equation}
P_g \mathcal{H}_{nn} P_g = -2 \gamma \sum_{k=0}^{d-1}  \eta_k^\dagger \eta_k \eta_{k+1}^\dagger  \eta_{k+1}.
\end{equation}
Contributions at second order of $\mathcal{H}_{nn}$ can be neglected because they depend on $\gamma^2/U \ll J^2/U$. Therefore the effective Hamiltonian for $U \gg J \gg \gamma$ is 
\begin{eqnarray}
\mathcal{H}_\text{eff} &=& - N \left( U + \frac{4 J^2}{U} \right) \sum_{k=0}^{d-1}  \eta_k^\dagger  \eta_k \nonumber \\ &-& \frac{2 J^2}{U} \sum_{k=0}^{d-1} ( \eta_k^\dagger \eta_{k+1} + h.c.) \nonumber \\ &-& \left (2\gamma - \frac{4 J^2}{U} \right) \sum_{k=0}^{d-1} \eta_k^\dagger \eta_k \eta_{k+1}^\dagger \eta_{k+1}.
\end{eqnarray}
Since for a fixed number of particles the first term is constant, we drop it in (\ref{Heff}).


\section*{Appendix C}

Recurrence equations (\ref{bs}) and (\ref{b1}) can be solved in basically the same way as (\ref{as}) and (\ref{a0}). Firstly, note that (\ref{bs}) is of the same form as (\ref{as}), therefore 
\begin{equation}
\beta_s = Br_0^{s} + Ar_0^{-s},
\end{equation}
where $A$ and $B$ are constants and
\begin{equation}
\epsilon=-2\bar{J}(r_0+r_0^{-1}).
\end{equation}
We assume $r_0 \leq 1$ and since $\lim_{s\rightarrow \infty} \beta_s = 0$ we get $A=0$. Therefore,
\begin{equation}
\beta_s=Br_0^s.
\end{equation}
We plug the above to (\ref{b1}) and obtain
\begin{equation}
(r_0+r_0^{-1}-\frac{\bar{\gamma}}{2\bar{J}})r_0 = r_0^2,
\end{equation}
which leads to 
\begin{equation}
r_0=\frac{2\bar{J}}{\bar{\gamma}} = \frac{\bar{J}}{\gamma-\bar{J}} = \frac{1}{\frac{\gamma U}{2J^2}-1}.
\end{equation}


\section*{Appendix D}

Consider the following state of $M$ fermionic A-B pairs represented by operators $\eta_k^{\dagger}=a_k^{\dagger} b_k^{\dagger}$
\begin{equation}
q_{(M)}^{\dagger N}|0\rangle,
\end{equation}
where 
\begin{equation}
q_{(M)}^{\dagger} = \frac{1}{\sqrt{d}}\sum_{k=0}^{d-1}\eta_{k}^{\dagger}\eta_{k+1}^{\dagger}\ldots\eta_{k+M-1}^{\dagger}.
\end{equation}
We are looking for the parameter
\begin{equation}
\chi_{N}^{(M)}=\frac{1}{N!}\langle 0|q_{(M)}^N q_{(M)}^{\dagger N}|0\rangle.
\end{equation}
Firstly, we write
\begin{equation}
\Gamma_k^{\dagger} = \eta_{k}^{\dagger}\eta_{k+1}^{\dagger}\ldots\eta_{k+M-1}^{\dagger}.
\end{equation}
The above operators obey
\begin{eqnarray}
\Gamma_k^{\dagger}\Gamma_{k'}^{\dagger}&=&\Gamma_{k'}^{\dagger}\Gamma_k^{\dagger}, \\ \Gamma_k^{\dagger}\Gamma_{k'}^{\dagger}&=&0~~\text{if}~~|k-k'|<M. \label{relationM}
\end{eqnarray}
We get
\begin{eqnarray}
q_{(M)}^{\dagger N}|0\rangle &=& \frac{1}{d^{N/2}}\sum_{k_1,\ldots,k_N}\Gamma_{k_1}^{\dagger}\ldots \Gamma_{k_N}^{\dagger}|0\rangle \nonumber \\
&=& \frac{N!}{d^{N/2}}\sum_{k_1 < \ldots < k_N}^{*}\Gamma_{k_1}^{\dagger}\ldots \Gamma_{k_N}^{\dagger}|0\rangle \nonumber \\
&\equiv & \frac{N!}{d^{N/2}}\sum_{k_1 < \ldots < k_N}^{*} |k_1,\ldots,k_N\rangle, \nonumber
\end{eqnarray}
where $*$ over the sum denotes that we take into account the relation (\ref{relationM}). This leads to
\begin{equation}\label{chiM}
\chi_N^{(M)} = \frac{N!}{d^N}\sum_{k_1 < \ldots < k_N}^{*} 1 = \frac{N!}{d^N}{{N+d-NM}\choose{d-NM}}.
\end{equation}
The binomial coefficient represents the combination with repetitions. The reason we get this value is the following. We consider $N$ compounds, each composed of $M$ A-B pairs. A single compound  occupies $M$ positions. In total there are $d$ positions, therefore after putting $N$ compounds there are $d-NM$ unoccupied spaces left. Each unoccupied space can be placed in one of $N+1$ possible places: before first compound, between the first and the second compounds, $\ldots$, after the $N$-th compound. Hence, we choose one of $d-NM$ positions from $N+1$ possibilities and the choices can repeat. In general, a choice of $x$ positions out of $y$, including repetitions, is given by ${x+y-1}\choose{x}$.    

\begin{figure}[H]
\includegraphics[width=0.5 \textwidth]{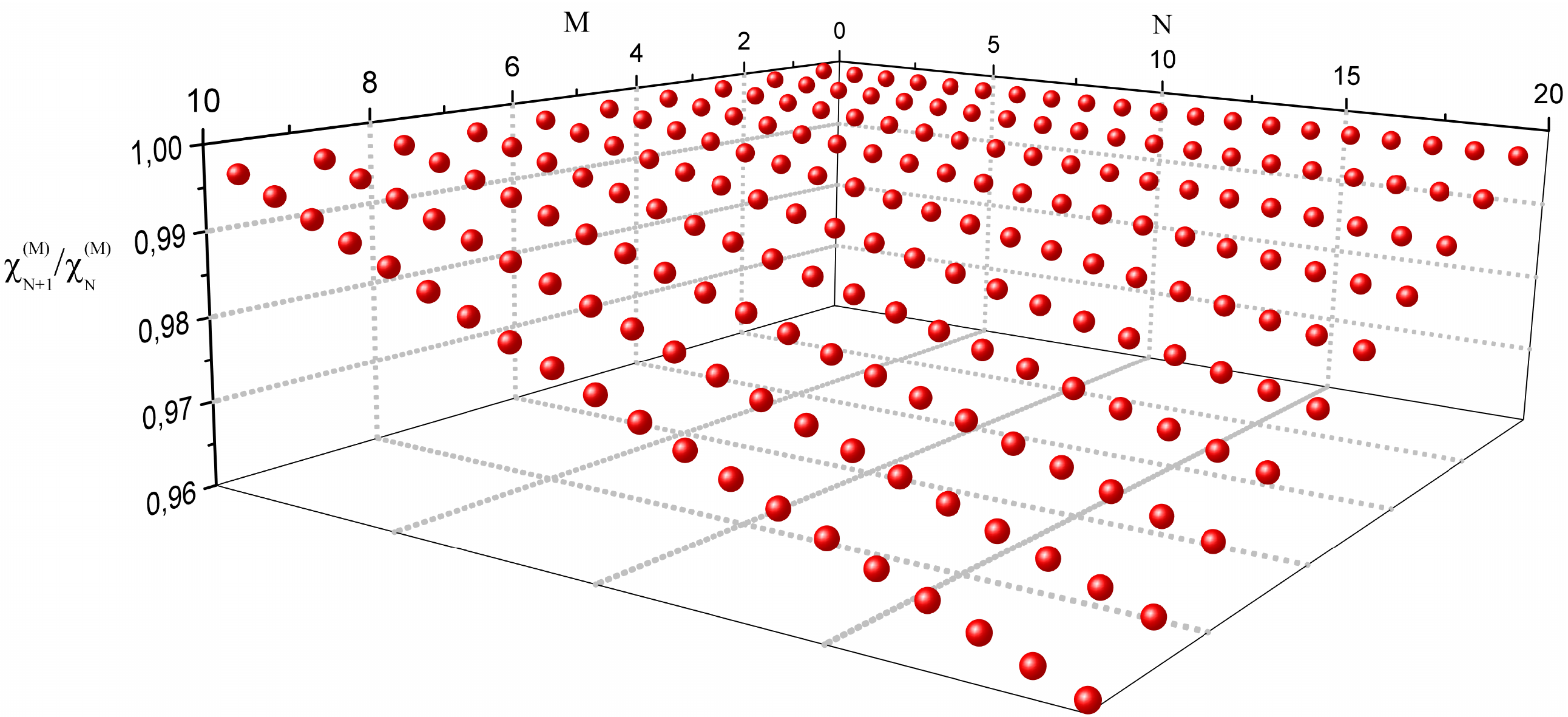}
\caption{The value of the lower bound of $\frac{\chi_{N+1}^{(M)}}{\chi_N^{(M)}}$ for $d=10000$. \label{figchi}}
\end{figure}

The formula (\ref{chiM}) can be written as
\begin{equation}
\chi_N^{(M)} = \frac{\Pi_{i=1}^{N}(d-NM+i)}{d^N}.
\end{equation}
Finally, we estimate the ratio $\frac{\chi_{N+1}^{(M)}}{\chi_N^{(M)}}$
\begin{eqnarray}
& &\frac{\chi_{N+1}^{(M)}}{\chi_N^{(M)}} = \frac{\Pi_{i=1}^{N+1}(d-(N+1)M+i)}{d\Pi_{i=1}^{N}(d-NM+i)}  \\
& & =\left(1-\frac{(N+1)(M-1)}{d}\right)\Pi_{i=1}^{N}\left(1-\frac{M}{d+i-NM}\right). \nonumber
\end{eqnarray}
The above value is upper-bounded by 1 and lower-bounded by
\begin{equation}
\left(1-\frac{(N+1)(M-1)}{d}\right)\left(1-\frac{M}{d+1-NM}\right)^N.
\end{equation}
In the limit $d\gg NM$ this value approaches 1. In Fig. \ref{figchi} we plot the above lower bound for $d=10000$.

\section*{Appendix E}

Let us show that a multifermionic state representing a single composite boson needs to be genuinely multipartite entangled. We analyse what happens if the state is not genuinely multipartite entangled. This means that it is separable with respect to some partition. In case of fermions separability implies that we can write it as
\begin{eqnarray}
c^{\dagger}|0\rangle &\equiv& \left(\sum_{i_1,i_2,\ldots}w_{i_1,i_2,\ldots} a^{\dagger}_{i_1}a^{\dagger}_{i_2}\ldots \right) \nonumber \\
&\times & \left(\sum_{j_1,j_2,\ldots}v_{j_1,j_2,\ldots} a^{\dagger}_{j_1}a^{\dagger}_{j_2}\ldots \right)|0\rangle,
\end{eqnarray}
where $w_{i_1,i_2,\ldots}$ and $v_{j_1,j_2,\ldots}$ are normalized antisymmetric coefficients corresponding to two partitions, such that 
\begin{equation}
\langle 0|cc^{\dagger}|0\rangle = \sum_{\substack{i_1,i_2,\ldots \\ j_1,j_2,\ldots }} |w_{i_1,i_2,\ldots}|^2|v_{j_1,j_2,\ldots}|^2= 1.
\end{equation}
For the moment we assume that these partitions can have an arbitrary number of fermions (here, without the loss of generality, we drop the division of fermions into A and B).

The main point of our argumentation is the following: if $c^{\dagger}|0\rangle$ represents a single composite boson, then $c^{\dagger 2}|0\rangle$ represents two composite bosons and its norm is approximately $2$. This is a necessary condition for bosonic creation and annihilation operators to obey the ladder structure (\ref{ladder}). We write explicitly
\begin{eqnarray}
c^{\dagger 2}|0\rangle &=& \sum_{\substack{i_1,i_2,\ldots \\ i'_1,i'_2,\ldots }} \left(w_{i_1,i_2,\ldots} a^{\dagger}_{i_1}a^{\dagger}_{i_2}\ldots \right)\left(w_{i'_1,i'_2,\ldots} a^{\dagger}_{i'_1}a^{\dagger}_{i'_2}\ldots \right) \nonumber \\
&\times & \sum_{\substack{j_1,j_2,\ldots \\ j'_1,j'_2,\ldots}} \left(v_{j_1,j_2,\ldots} a^{\dagger}_{j_1}a^{\dagger}_{j_2}\ldots \right) \left(v_{j'_1,j'_2,\ldots} a^{\dagger}_{j'_1}a^{\dagger}_{j'_2}\ldots \right)|0\rangle. \nonumber
\end{eqnarray}
For non-overlapping terms (the ones containing different fermionic creation operators) we have 
\begin{eqnarray}
\left(w_{i_1,i_2,\ldots} a^{\dagger}_{i_1}a^{\dagger}_{i_2}\ldots \right)\left(w_{i'_1,i'_2,\ldots}a^{\dagger}_{i'_1}a^{\dagger}_{i'_2}\ldots \right) = \nonumber \\
x\left(w_{i'_1,i'_2,\ldots}a^{\dagger}_{i'_1}a^{\dagger}_{i'_2}\ldots \right)\left(w_{i_1,i_2,\ldots} a^{\dagger}_{i_1}a^{\dagger}_{i_2}\ldots \right) \label{twicecount}
\end{eqnarray} 
(same holds for the terms in the other partition). In the above $x=+1$ ($x=-1$) if the number of fermions in each partition is even (odd). If $x=-1$ all the terms cancel and $c^{\dagger 2}=0$. Therefore, we allow only for partitions with even number of fermions. We can write
\begin{eqnarray}
c^{\dagger 2}|0\rangle &=& 4\sum_{\substack{i_1,i_2,\ldots \\ i'_1,i'_2,\ldots \\ i_1 < i'_1 }}^{\ast} \left(w_{i_1,i_2,\ldots} w_{i'_1,i'_2,\ldots} a^{\dagger}_{i_1}a^{\dagger}_{i_2}\ldots a^{\dagger}_{i'_1}a^{\dagger}_{i'_2}\ldots \right) \nonumber \\
&\times & \sum_{\substack{j_1,j_2,\ldots \\ j'_1,j'_2,\ldots \\ j_1 < j'_1}}^{\ast} \left(v_{j_1,j_2,\ldots}v_{j'_1,j'_2,\ldots} a^{\dagger}_{j_1}a^{\dagger}_{j_2}\ldots  a^{\dagger}_{j'_1}a^{\dagger}_{j'_2}\ldots \right)|0\rangle. \nonumber
\end{eqnarray}
The factor of 4 stems from (\ref{twicecount}), i.e.,  the terms corresponding to products within each partition are counted twice. If there were $s$ partitions, the overall factor would be $2^s$. In addition, the symbol $\ast$ above the sums indicates that we do not sum over the terms with overlapping indices. For example, due to Pauli exclusion we do not sum over the terms corresponding to $w_{1,4,7,\ldots}w_{2,4,9,\ldots}$ etc.   

We obtain
\begin{eqnarray}
\langle 0|c^2 c^{\dagger 2}|0\rangle &=& 16 \sum_{\substack{i_1,i_2,\ldots \\ i'_1,i'_2,\ldots \\ i_1 < i'_1 }}^{\ast}\sum_{\substack{j_1,j_2,\ldots \\ j'_1,j'_2,\ldots \\ j_1 < j'_1}}^{\ast} \left( |w_{i_1,i_2,\ldots}|^2 \right. \nonumber  \\ &\times & \left. |v_{j_1,j_2,\ldots}|^2|w_{i'_1,i'_2,\ldots}|^2|v_{j'_1,j'_2,\ldots}|^2 \right).
\end{eqnarray}  
The above can be rewritten as
\begin{eqnarray}
& &4\sum_{\substack{i_1,i_2,\ldots \\ i'_1,i'_2,\ldots }}\sum_{\substack{j_1,j_2,\ldots \\ j'_1,j'_2,\ldots }} |w_{i_1,i_2,\ldots}|^2 |v_{j_1,j_2,\ldots}|^2|w_{i'_1,i'_2,\ldots}|^2|v_{j'_1,j'_2,\ldots}|^2  \nonumber \\
&-&4\omega(\ast) = 4(1-\omega(\ast)),
\end{eqnarray} 
where 
\begin{equation}
\omega(\ast)= \sum^{\ast}|w_{i_1,i_2,\ldots}|^2 |v_{j_1,j_2,\ldots}|^2|w_{i'_1,i'_2,\ldots}|^2|v_{j'_1,j'_2,\ldots}|^2
\end{equation}
and the sum is taken only over terms with overlapping indices. If the number of modes that can be occupied by fermions is arbitrarily large and the states of each partition are highly entangled (e.g., they are the ones discussed in Appendix D), then the parameter $\omega(\ast)$ can be arbitrarily small. Therefore, the norm of $c^{\dagger 2}|0\rangle$ is approximately four, not two. If instead of two partitions we considered $s$ partitions, the norm would be approximately $2^s$. For a proper bosonic behaviour we need $s=1$, which implies genuinely multipartite entangled state. 

Finally, we need to make one remark. The norm of $c^{\dagger 2}|0\rangle$ can be in principle equal to two, even for $s=2$. This requires $\omega(\ast)=1/2$, i.e., a large number of terms with overlapping indices (occurring for not that strongly entangled states). However, the operator $c^{\dagger}$ also needs to recover the ladder structure for higher powers than two. If $\omega(\ast)=1/2$, then $\langle 0|c^N c^{\dagger N}|0\rangle \neq N!$ for $N>2$. In fact, for a large number of overlapping terms we get $\langle 0|c^N c^{\dagger N}|0\rangle \rightarrow 0$ for relatively small N. This is the consequence of Pauli exclusion.




\begin{thebibliography}{99}
	\bibitem{Law}
	C. K. Law, Phys. Rev. A. {\bf 71}, 034306 (March 2005).
	
	\bibitem{Wootters}
	C. Chudzicki, O. Oke, and W. K. Wootters, Phys. Rev. Lett. {\bf 104}, 070402 (2010).

	\bibitem{Bouvrie2012a}
	M. C. Tichy, P. A. Bouvrie, and K. M\o{}lmer, Phys. Rev. A \textbf{86}, 042317 (2012).
	
	\bibitem{Bouvrie2012b}
	M. C. Tichy, P. A. Bouvrie, and K. M\o{}lmer, Phys. Rev. Lett. \textbf{109}, 260403 (2012).
		
	\bibitem{Bouvrie2013}
	M. C. Tichy, P. A. Bouvrie, and K. M\o{}lmer, Phys. Rev. A \textbf{88}, 061602 (2013).
	
	\bibitem{Bouvrie2014}
	M. C. Tichy, P. A. Bouvrie, and K. M\o{}lmer, Appl. Phys. B \textbf{117}, 785 (2014).

	\bibitem{Bouvrie2016}
	P. A. Bouvrie, M. C. Tichy, and K. M\o{}lmer, Phys. Rev. A \textbf{94}, 053624 (2016).
	
	\bibitem{Bouvrie2017}
	P. A. Bouvrie, M. C. Tichy, and I. Roditi, Phys. Rev. A \textbf{95}, 023617 – (2017).
	
	\bibitem{Bouvrie2018}
	P. A. Bouvrie, E. Cuestas, I. Roditi and A. P. Majtey,	arXiv:1810.07827 (2018).
	
	\bibitem{pawel2011}
	R. Ramanathan, P. Kurzy\'{n}ski, T. K. Chuan, M. F. Santos, and D. Kaszlikowski, Phys. Rev. A \textbf{84}, 034304 (2011).
		
	\bibitem{Pawel2012}
	P. Kurzy\'{n}ski, R. Ramanathan, A. Soeda, T. K. Chuan, and D. Kaszlikowski, New J. Phys. \textbf{14}, 093047 (2012).
	
	\bibitem{Pawel2013}
	S.-Y. Lee, J. Thompson, P. Kurzy\'{n}ski, A. Soeda, and D. Kaszlikowski, Phys. Rev. A \textbf{88}, 063602 (2013).

	\bibitem{Pawel2015}
	S.-Y. Lee, J. Thompson, S. Raeisi, P. Kurzy\'{n}ski, and D. Kaszlikowski, New J. Phys. \textbf{17}, 113015 (2015).
			
	\bibitem{zak2017}
	Z. Lasmar, D. Kaszlikowski, and P. Kurzy\'{n}ski, Phys. Rev. A {\bf96}, 032325 (2017).
		
	\bibitem{zak2018}
	Z. Lasmar, A. S. Sajna, S.-Y. Lee, and P. Kurzynski, Phys. Rev. A \textbf{98}, 062105 (2018).
	
		\bibitem{Fried1998}
	D. G. Fried, T. C. Killian, L. Willmann, D. Landhuis, S. C. Moss, D. Kleppner, and T. J. Greytak. Phys. Rev. Lett. 
	\textbf{81}, 3811 (1998).
	
	\bibitem{Silvera1980}
	I. F. Silvera and J. T. M. Walraven	Phys. Rev. Lett. \textbf{44}, 164 (1980).

	\bibitem{Slater1}	
	J. Schliemann, J. I. Cirac, M. Ku ́s, M. Lewenstein, and D. Loss, Phys. Rev. A {\bf 64}, 022303 (2001).
	
	\bibitem{Slater2}
	Y. S. Li, B. Zeng, X. S. Liu, and G. L. Long, Phys. Rev. A {\bf 64}, 054302 (2001).
	
	\bibitem{Slater3} 
	R. Paskauskas and L. You, Phys. Rev. A {\bf 64}, 042310 (2001).
		
	\bibitem{Eckert2002}
	K. Eckert, J. Schliemann, D. Bru{\ss}, M.Lewenstein, Annals of Physics, \textbf{299} (1), 88 (2002).	
		
	\bibitem{Monique2001}
	M. Combescot and C. Tanguy. Europhys. Lett., \textbf{55} (3), 390 (2001).
	
	\bibitem{Monique2003a}
	M. Combescot and C. Tanguy, Eur. Phys. J. B 31, 17–24 (2003).
	
	\bibitem{Monique2003b}
	S. Rombouts , D. Van Neck and L. Pollet, Europhys. Lett., \textbf{63} (5), 785 (2003)	
	
	\bibitem{Monique2008}
	M. Combescot, O. Betbeder-Matibet, and F. Dubin, Phys. Rep. {\bf 463}, 215 (2008).
	
	\bibitem{Monique2009}
	M. Combescot, F. Dubin, andM. A. Dupertuis, Phys. Rev. A 80, 013612 (2009).
	
	\bibitem{Monique2010}
	M. Combescot and O. Betbeder-Matibet, Phys. Rev. Lett. 104, 206404 (2010).
	
	\bibitem{Monique2011a}
	M. Combescot, S.-Y. Shiau, and Y.-C. Chang, Phys. Rev. Lett. 106, 206403 (2011).
	
	\bibitem{Monique2011b}
	M. Combescot, Europhys. Lett. 96, 60002 (2011). 

	\bibitem{Monique2015}
	M. Combescot, R. Combescot, M. Alloing, and F. Dubin, Phys. Rev. Lett. \textbf{114}, 090401 (2015).
	
	\bibitem{Monique2016}
	M. Combescot, S.-Y. Shiau, and Y.-C. Chang, Phys. Rev. A \textbf{93}, 013624 (2016).

	\bibitem{Thilagam2013}
	A. Thilagam, J. Math. Chem. \textbf{51}, 1897 (2013).
		
	\bibitem{Thilagam2015}
	A. Thilagam, Physica B: Condensed Matter \textbf{457}, 232 (2015).
	
	\bibitem{ShiJian2004}
	S.-J. Gu, S.-S. Deng, Y.-Q. Li, and H.-Q. Lin, Phys. Rev. Lett. \textbf{93}, 086402 (2004).
	
	\bibitem{BouvrieValdesetall2016}
	P. A. Bouvrie, A. Valdes-Hernandez, A. P. Majtey, C. Zander, and A. R. Plastino, Annals of Physics \textbf{383}, 401 (2017).
	
	\bibitem{cespedes}
	P. Cespedes, An\'alisis de la validez de la teor\'ia de cobosones en un modelo simple, Master's thesis, FaMAF, Universidad Nacional de C\'ordoba, 
	
	\bibitem{Lieb1968}
	E. H. Lieb, F. Y. Wu, Phys. Rev. Lett. \textbf{20}, 1445 (1968).
	
	
	
	
\end{thebibliography}
\end{document}